\documentclass[12pt,a4paper]{article}
\usepackage{amssymb,amsfonts,times,graphicx,psfrag, epsf}
\makeatletter

\@addtoreset{equation}{section}
\makeatother

 \def\lmu{ \< \!\mu \left|\, } \def\rmu{\, \right|\mu\!  \>} 
\newcommand{\nn}{\nonumber}
\newcommand{\vev}[1]{\left\langle{#1}\right\rangle}
\newcommand{\tfrac}[2]{{\textstyle\frac{#1}{#2}}}

\newcommand{\bS}{{\bf S}}
\newcommand{\ZZ}{{\mathbb Z}}

\newcommand{\RR}{{\mathbb R}}

\def\e{\epsilon}
\def\hf{ {\textstyle{\frac12}} }
\def\({\left(}
\def\){\right)}
\def\<{\left\langle\,}
\def\>{\, \right\rangle}
\def\a{\alpha}

\newcommand{\eqal}[2]{\begin{eqnarray} #2 \end{eqnarray}}

\newcommand{\eqn}[2]{\begin{equation} #2 \end{equation}}

\def\bea{\begin{eqnarray}}
\def\eea{\end{eqnarray}}
\def\nn{\nonumber}
\def\beq{\begin{equation}}
\def\eeq{\end{equation}}
\def\be{\bea\new\begin{array}{c}}  
\def\ee{\end{array}\eea}           
\def\bse{\begin{subequations}}                
\def\ese{\end{subequations}}

  \def\p{\partial}
     \def\eh{\hat{\epsilon}} 
 \def\Wp{W\!_{\!  _+}} 
 \def\Wm{W\!_{\!  _-}}
  \def\Wpm{W\!_{\!  _ \pm}}

 \def\tr{ {\rm tr\, } } 
\def\la{ \label} 
\def\xp{x^{^{+}}} 
\def\xm{x^{^{-}}}

\def\xip{\xi^{_{+}}} 
\def\xim{\xi^{_{-}}} 
\def\xipm{\xi^{_{\pm}}} 
 
\def\CS{{\cal S}}
\def\CR{{\cal R}}
\def\O{\Omega}

 \def\CD{{\cal D}}
\def\s{\sigma}
 \def\x{\xi}
\def\d{\delta}
\def\kk{ E}
 \def\tL{{\tilde  \Lambda}}
 
\def\muB{\mu_{_{B}} }
\def\sIR{  L}
\newcommand{\cH}{\mathcal{H}}
\newcommand{\cV}{{\mathcal{V}}}
\newcommand{\hh}{\hat{h}}
\newcommand{\rE}{\overrightarrow{E}}
\newcommand{\lE}{\overleftarrow{E}}
\newcommand{\X}{\mathbf{X}}
\def\Xp{\X_{_{+}}} 
\def\Xm{\X_{_{-}}} 
\def\Xpm{\X_{_{\pm}}}

\begin{document}
\thispagestyle{empty}
\begin{flushright}
{SPhT-T07/071}\\
{UT-07-29}
\end{flushright}
\vspace{1cm}
\setcounter{footnote}{0}
\begin{center}
{\Large{\bf Scattering of Long Folded Strings and Mixed Correlators in
the Two-Matrix Model}}

\vspace{20mm} {\sc J.-E. Bourgine$^{\ast }$}, {\sc K. Hosomichi$^{\ast
}$}, {\sc I. Kostov}$^{\ast \circ}$ and {\sc Y. Matsuo}$^{\dagger}$
\\[7mm] {\it $^\ast$Service de Physique Th\'eorique, CNRS-URA 2306 \\
C.E.A.-Saclay \\
F-91191 Gif-sur-Yvette, France}\\[1mm]

{\it $^\dagger $ {\it Department of Physics, Faculty of Science,
University of Tokyo \\
Hongo 7-3-1, Bunkyo-ku, Tokyo 113-0033, Japan\\} }

\vskip 2cm

{\sc Abstract}\\[2mm]
\end{center}
 
\noindent{ We study the interactions of Maldacena's long folded
strings in two-dimensional string theory.  We find the amplitude for a
state containing two long folded strings to come and go back to
infinity.  We calculate this amplitude both in the worldsheet theory
and in the dual matrix model, the Matrix Quantum Mechanics.  The
matrix model description allows to evaluate the amplitudes
involving any number of long strings, which are given by the mixed
trace correlators in an effective two-matrix model.
}
 
\vfill

{\small $^\circ$Associate member of the {\it Institute for Nuclear
Research and Nuclear Energy, Bulgarian Academy of Sciences, 72
Tsarigradsko Chauss\'ee, 1784 Sofia, Bulgaria}}

\newpage

\setcounter{page}{1}

\section{Introduction}
\label{sec:intro}

In this paper we study the scattering of long folded strings in 2D
string theory.  As pointed out by Maldacena \cite{Malong}, long
folded strings stretching from infinity correspond to non-singlet
states in the dual matrix model.  It is expected that condensation of
such states can produce curved background with horizon 
\cite{FZZconj, KKK}.
Our main motivation for this work is to study the possibility of
formulating a Lorentzian version of the ``black hole matrix model",
discussed in \cite{Malong}.  Our results suggest that the chiral
formalism introduced in \cite{AKK} and further developed in
\cite{flows, streq, AKKnorm, ADKMV, Yin, Tesc, Mukh} is well adapted
for this purpose.
 
\smallskip

The two-dimensional strings have only longitudinal modes, and the
closed string spectrum is that of a single massless particle, the
`tachyon' \cite{reviews}.  In addition to the closed string spectrum,
the theory has states of infinite energy, associated with long folded
strings stretched to infinity in the space direction $\phi$.  Folded
strings in two dimensions were studied in \cite{Bars:1994sv} and more
recently in \cite{Malong, Gaiotto, Seib}.  Such strings have infinite
energy since they stretch all the way to $\phi\to- \infty$.  After
subtracting the infinite part, the spectrum is unbounded from below.
Any physical observable in such a theory can be formulated as a
scattering amplitude relating incoming right moving and outgoing left
moving states.  The asymptotic states can be thought of as composed of
quasiparticles.  Each such quasiparticle represents the tip of a
folded string.

An exact worldsheet description of folded strings based on Liouville
string theory was given by Maldacena \cite{Malong}.  He argued that a
stack of FZZT branes placed far away in the asymptotically free region
($\muB \gg\sqrt\mu$) can be considered as a source for long folded
strings.  The evolution of a long string starts with a very energetic
short open string in the region $\phi\ll - \log\muB$.  When the ends
of the string reach $\phi \sim - \log\muB$, they get trapped by the
brane, while the bulk of the string continues to move until it looses
all its kinetic energy at distance $\phi\sim -\log\mu$ and starts to
evolve back.  This picture allows to express the reflection amplitude
for the tip of a long folded string as a certain limit of the boundary
two-point function in Liouville theory.  Using the expression for this
correlation function found in \cite{FZZb}, Maldacena gave an explicit
formula for this reflection amplitude.
 \smallskip
 
In the dual matrix model, the Matrix Quantum Mechanics (MQM), the
closed strings propagate in the singlet sector, while the folded
strings propagate in the non-singlet sector of MQM, characterized by
the presence of Wilson lines.  In MQM, the states containing one
folded string are those in the adjoint representation.  They can be
considered as impurities in the fermi sea.  The wave function of such
states depends on a collective coordinate giving the position of the
tip of the folded string.  It satisfies a Calogero type equation,
whose collective field formulation was given in \cite{Malong}.  The
explicit solution of this equation was found later in \cite{Fid}, and
the result for the scattering phase was identical with the one
obtained from the worldsheet theory.

 The states with $n$ folded strings, or $n$ impurities, are described
 by irreducible representations whose Young tableaux contain $n$ boxes
 and $n$ anti-boxes.  These are the representations that occur in the
 direct product of $n$ fundamental and $n$ anti-fundamental
 representations.

 The extension of the canonical formalism of MQM to higher
 representations passes through the solution of the corresponding
 Calogero problem, which seems to be a quite difficult, although not
 impossible, task.  Instead one can try to attack the problem using
 the chiral quantization of MQM, which operates directly in terms of
 asymptotic incoming and outgoing states.  Here the Hamiltonian is
 first order and therefore has no Calogero term.  Using the chiral
 formalism, the scattering problem in the non-singlet sector of MQM
 was reformulated by one of the authors \cite{Iadj} in terms of the
 mixed trace correlators in an effective two-matrix model.  This
 allowed to apply some powerful results derived for the two-matrix
 model \cite{ EynM,  EO}.  In particular, it was shown in
 \cite{Iadj} that the scattering amplitude in the adjoint
 representation, evaluated originally in \cite{Malong, Fid}, coincides
 with the simplest mixed trace correlator in the effective two-matrix
 model.

In this paper we evaluate, using the chiral formalism of MQM, the
reflection amplitudes of higher non-singlets, focusing mainly on the
case $n=2$.  The reflection amplitude can be expanded in the inverse
cosmological constant $g_{\rm s}\sim 1/\mu$.  The leading term is the
Young-symmetrized product of the reflection amplitudes for two
non-interacting quasiparticles.  The interaction appears in the
subleading term, for which we find an explicit expression.

We give two independent derivations of the subleading term, performed
in the worldsheet theory and in the matrix model.  In the worldsheet
theory, the subleading term is given by the 4-point boundary amplitude
in a suitable limit.  In the derivation we make a heavy use of the
symmetries imposed by the boundary ground ring.  To set the notations
and explain the problem, we first present the derivation of the $n=1$
amplitude, the reflection factor for a single long string, originally
obtained in \cite{Malong}.
  
The result is unexpectedly simple.  We find that the reflection
amplitude in the subleading order consists of two terms, which have a
natural interpretation in terms of reflection and scattering of the
two quasiparticles.  The first term describes a scattering of the two
quasiparticles with non-zero energy transfer, followed by reflection
of each quasiparticle.  In the second term the scattering and the
reflections occur in the opposite order.  The scattering amplitude for
two quasiparticles does not depend on $\mu$ and $\muB$ and therefore
occurs in the extreme asymptotic domain $\phi\ll -\log\muB$, where the
incoming and the outgoing strings are short.

The matrix model description allows to evaluate the amplitudes for
states with any number $n$ of quasiparticles.  We first evaluate the
amplitudes in the coordinate space and then perform a Fourier
transformation.  We performed explicitly the Fourier
transformation for the case $n=2$ and reproduced the result of the
worldsheet theory.  We observed that the $n=2$ amplitudes in the
coordinate and momentum space essentially coincide.  Assuming that
this is a general property, we speculate about the structure of the
reflection amplitude for states containing $n$ long strings having a
common worldsheet with the topology of a disk.  We argue that such an
amplitude again decomposes into elementary processes, reflections of
long strings and scattering of any number $k\le n$ of short open
strings.

\section{Long folded strings in worldsheet theory}

The worldsheet theory is described by a free boson $X$ and a
Liouville field $\phi$.  The field $X$ is regarded as time, so it has
the opposite signature.  As was done in \cite{DiFrancesco:1991ud}, it
is convenient to consider a family of theories in which both $\phi$
and $X$ couple to the worldsheet curvature.  Their background charges
are $Q=b+1/b$ and $\tilde Q=ib-i/b$ respectively, so that matter
central charge is critical,
\[
 c_{\rm matter} = 2+6(Q^2+\tilde Q^2)=26.
\]
We will be mostly interested in the case $b=1$, but will keep $b$ as
arbitrary until the final stage in order to avoid singularities which
are peculiar to $b=1$.

Local bulk operators 
$$
T^\pm_k\sim e^{(Q\pm ik)\phi-i(\tilde Q+k)X}
$$ 
of marginal dimension correspond to on-shell tachyon modes with energy
$k$.  They are right or left-moving waves depending on the sign
choice.  Similarly, local boundary operators
$$
U^\pm_k\sim e^{(\frac
Q2\pm ik)\phi-i(\frac{\tilde Q}2+k)X}
$$ 
correspond to physical open string modes with energy $k$.  In this
paper we focus on the open strings ending on FZZT-branes.

The action for $\phi$ has a potential $\mu e^{2b\phi}$ which scatters
every incoming (right-moving) tachyon back to $\phi=-\infty$.  As a
consequence, the operators $T_k^+$ and $T_k^-$ are proportional to
each other.  A similar relation holds also for the open string
operators $U_k^+$ and $U_k^-$, but the relation becomes more
complicated because the end-points of the open strings also feel the
boundary potential $\muB e^{b\phi}$.  We label the branes by $s$, in
terms of which $\muB $ can be expressed as
\[
 \muB (s) ~=~ \sqrt{\frac{\mu}{|\sin\pi b^2|}}\cosh(2\pi bs).
\]
We consider FZZT-branes with very large $s$.  If one throws in an open
string ending on such branes, its endpoints first reach the boundary
potential wall at $\phi\sim -2\pi s-\frac{1}{2b}\log\mu$, which is
much before the bulk potential wall at $\phi\sim
-\frac{1}{2b}\log\mu$.  When the endpoints are caught by the
potential, the string starts to stretch and its tip continues to move
towards the strong coupling region until it uses up all its kinetic
energy.  The tip of such a string can probe the bulk Liouville wall if
it initially has a sufficiently large energy, $k\sim 2s$.  This is how
a long folded string is realized in two-dimensional string theory
\cite{Malong}.

\subsection{Classical analysis}

\def\ssp  {\sigma_{\! \!_{ +}}}
\def\ssm{\sigma_{\! \!_{ -}}}
\def\sspm{\sigma_{\! \!_{ \pm}}}

\def\sp  {s_{\! _{ +}}}
\def\sm{s_{\! _{ -}}}
\def\spm{s_{\! _{ \pm}}}

The classical motion of an open string is described by the action
\begin{eqnarray}
 S &=& \int\frac{d\tau d\sigma}{4\pi}
 \left\{(\partial_\tau\phi)^2-(\partial_\sigma\phi)^2
       -(\partial_\tau X  )^2+(\partial_\sigma X)^2
       -\mu\pi e^{2b\phi} \right\}
 \nn\\ && 
 -\int d\tau \{\muB (\sp )e^{b\phi(\ssp  ,\tau)}
              +\muB (\sm )e^{b\phi(\ssm ,\tau)}\},
\label{Scl}
\end{eqnarray}
defined on a strip $\sigma\in [\ssm ,\ssp ]$, $\tau\in\RR$.  The
classical argument is known to be valid for small $b$.  We focus on
the solutions parametrized by $\gamma$,
\begin{equation}
 X ~=~ \tau,~~~~ 4\mu\pi e^{2b\phi} ~=~ \left\{\cosh b\gamma\cosh
 b\tau+\sinh b\gamma\cosh b\sigma \right\}^{-2}\, ,
\label{solXphi}
\end{equation}
which solve the bulk equation of motion as well as the Virasoro
constraint $T_{\pm\pm}=0$.  These solutions were first presented in
the context of long folded strings in \cite{Malong}.  The tip of the
folded string is at $\sigma=0$, and $\phi$ reaches maximum at
$\sigma=\tau=0$.

\begin{figure}[htb]
{
\centerline{
\scalebox{1.2}{
\scriptsize
\psfrag{phi}{$\phi$}
\psfrag{wall}{$~~~~0$}
\psfrag{bwall}{$-2\pi s$}
\psfrag{sigma}{$\sigma$}
\psfrag{tau}{$X=\tau$}
\includegraphics{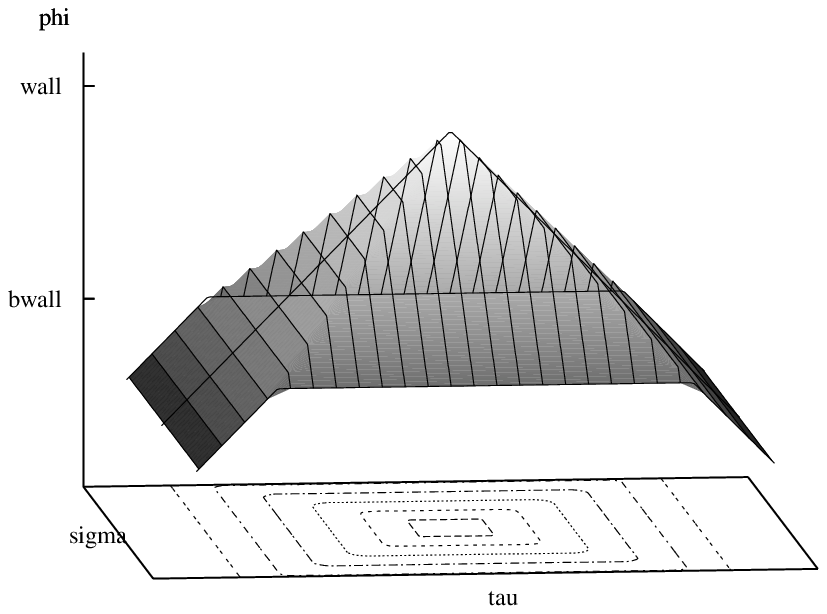}
}~~~~~
\scalebox{0.7}{
\psfrag{(A)}{$(A)$}
\psfrag{(B)}{$(B)$}
\psfrag{(C)}{\hskip-3mm$(C)$}
\psfrag{(D)}{$(D)$}
\psfrag{tt}{$\tau$}
\psfrag{ss}{$\sigma$}
\includegraphics{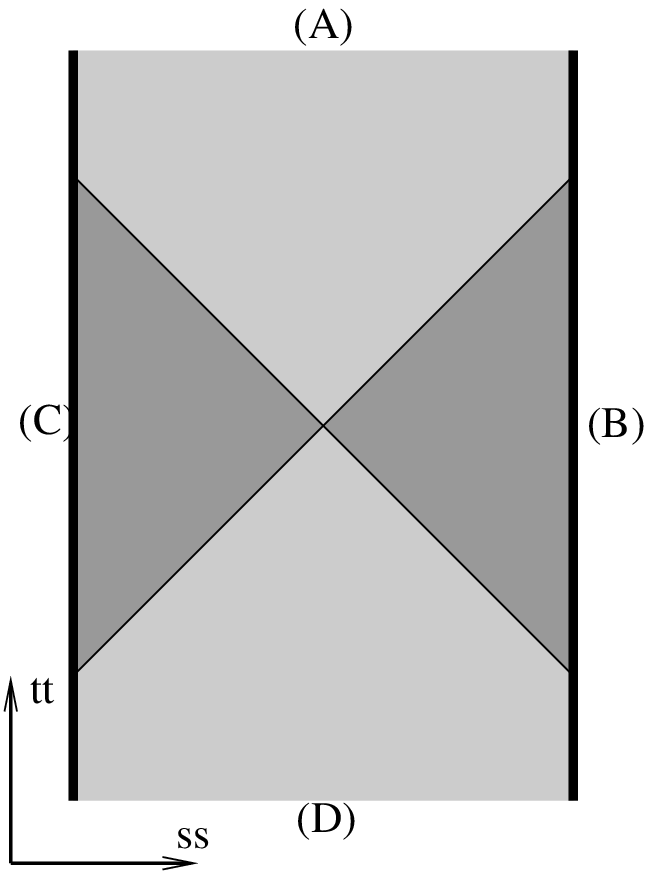}}
  } } 
  \caption{ \small{(Left) A classical long folded string for
  $\mu=b=1$, $\gamma=10$, $\sspm =\pm500$.  Some equal-$\phi$ lines
  are drawn on the base $\sigma$-$\tau$ plane.  (Right) The string
  worldsheet is divided into four regions according to the behavior of
  $\phi$.  }}
\label{fig:classical}
\end{figure}

The boundary conditions on fields, $\partial_\sigma X =
\partial_\sigma\phi \pm 2\pi b\muB (\spm ) e^{b\phi} = 0$, are
satisfied if
\begin{equation}
\sinh^2 b\gamma\sinh^2 b\sspm  ~=~ \cosh^2(2\pi b\spm ).
\end{equation}
These relations allow us to express $(k,\spm )$ as functions of
$(\gamma,\sspm )$.  For very large $\spm $ and a finite $\gamma$,
$\sspm $ roughly equals $2\pi \spm $ up to sign.  Long folded open
strings correspond to the choice $\ssm <0<\ssp $.  The (spacetime)
energy $k$ of such a string is given by
\begin{equation}
2\pi k  =(\ssp   -\ssm  )/ 2\pi   \sim \sp  + \sm .
\end{equation}
See the Figure \ref{fig:classical} for an example of a long folded
string.  The other two choices, $0<\ssm <\ssp $ or $\ssm <\ssp <0$,
both lead to ``short'' strings which do not develop long folded
worldsheets.  The role of short strings is important in understanding
the interactions of folded long strings.  \vskip 10pt

\paragraph{Reflection amplitude.}

In the tree approximation, the phase for reflection amplitude of a
long folded string is given by the classical value of the action
(\ref{Scl}) on the strip.  Let us give a rough evaluation of it for
the solution presented above, assuming that $\gamma$ is reasonably
large.  The worldsheet strip is decomposed into four regions according
to the behavior of $\phi(\tau,\sigma)$:
\begin{eqnarray}
 (A) ~:~ \tau>+|\sigma| && \phi ~\sim~ \phi_0-\tau, \nn\\
 (B) ~:~ \sigma>+|\tau| && \phi ~\sim~ \phi_0-\sigma, \nn\\
 (C) ~:~ \sigma<-|\tau| && \phi ~\sim~ \phi_0+\sigma, \nn\\
 (D) ~:~ \tau<-|\sigma| && \phi ~\sim~ \phi_0+\tau,
\la{clls}
\end{eqnarray}
where $\phi_0=-\gamma-\frac{1}{2b}\log(\frac{\mu\pi}4)$ is a constant.
See the right of the Figure \ref{fig:classical}.  The bulk and
boundary potential terms in the action can be neglected in the limit
of large $s_\pm$.  The contributions to the classical action from the
kinetic terms of $\phi$ and $X$ cancel in the regions $(A)$ and $(D)$,
whereas they add up in the regions $(B)$ and $(C)$.  The classical
action is therefore roughly proportional to the areas of the regions
$(B)$ and $(C)$:
\begin{equation}
 S_{\rm cl} ~\sim~ - \frac{1}{2\pi}(\ssp ^2 +\ssm ^2) ~\sim~ -
 2\pi(\sp ^2 +\sm ^2).
\label{SclL}
\end{equation}

\subsection{Quantum theory}

\def\nuu{\nu } We consider the following physical (on-shell) boundary
operators
\begin{eqnarray}
 U_k^+ &=& b^{\frac12}\Gamma(-2ikb)\, \nuu  ^{\frac Q2+ik}
           \, {\rm c} \, e^{(\frac Q2+ik)\phi-i(\frac{\tilde Q}2+k)X},
 \nn\\
 U_k^- &=& b^{-\frac12}\Gamma(2ik/b)\, \nuu  ^{\frac Q2-ik}
   \,       {\rm c} \, e^{(\frac Q2-ik)\phi-i(\frac{\tilde Q}2+k)X},
   \label{vertop}
\end{eqnarray}
corresponding to right- or left-moving open string excitations.  Here
${\rm c}$ is the reparametrization ghost and we have introduced
$$\nuu   \equiv\{\mu\pi\gamma(b^2)\}^{1/2b},
 \qquad \gamma(x)\equiv\frac{\Gamma(x)}{\Gamma(1-x)}.$$
   The operators $U_k^+$ and
$U_k^-$ are proportional to each other,
\[
 ^s[U_k^+]^{s'}~=~ d(k,s,s')\cdot{}^s[U_k^-]^{s'},
\]
where one should remember that the properties of boundary operators
depend also on the two D-branes.  The proportionality constant
$d(k,s,s')$ is essentially given by the Liouville boundary reflection
coefficient $d_L(\beta,s,s')$ \cite{FZZb}:
\begin{eqnarray}
 d(k,s,s') &=&
 b\, \nuu  ^{2ik}\, \frac{\Gamma(-2ikb)}{\Gamma(2ik/b)}\, 
 d_L(\tfrac Q2+ik,s,s')
 \nn\\ &=&
 \bS(2ik+\tfrac1b)\, 
 \bS(\tfrac Q2-i(k+s+s'))\, \bS(\tfrac Q2-i(k+s-s'))
 \nn\\ &&\hskip15mm \times\, 
 \bS(\tfrac Q2-i(k-s+s'))\, \bS(\tfrac Q2-i(k-s-s')).
\label{dkss}
\end{eqnarray}
The function $\bS(x)$ is introduced and used in \cite{FZZb}.  Some of
its properties are collected in Appendix A.

We will consider Maldacena's limit \cite{Malong} in which all boundary
parameters have the form
$$s=\sIR +\delta s$$
with $\sIR$ assumed to be very large, while $\delta s$ is kept finite.
The FZZT-branes of our interest are all labeled by such $s$.  A folded
long open string is described by a pair of vertex operators $U^+_{k}$
and $U^-_{-k}$ with
$$k\sim 2\sIR.$$
We also consider short open strings carrying finite energy $k$, as
they will appear as intermediate particles in the scattering of long
folded strings.  The reflection amplitude of a single open string is
given by $d(k,s,s')$.  Using the asymptotics of $\bS(x)$ one finds
\begin{eqnarray}
  d(k,s,s')|_{(k,s,s')\sim(0,\sIR ,\sIR )}~~~
  &=& e^{-4\pi i\sIR k+{\cal O}(\sIR ^0)},
  \nn\\
  d(k,s,s')|_{(k,s,s')\sim(2\sIR ,\sIR ,\sIR )}
  &=&e^{-2\pi i(s^2+s'^2)+2\pi i\sIR \tilde Q+{\cal O}(\sIR ^0)},
\label{asympd}
\end{eqnarray}
for short and long open strings respectively.  To the leading order in
$\sIR$, the second formula agrees with the classical result
(\ref{SclL}) for long folded strings.  The full expression for the
reflection amplitude at $b=1$ is
\begin{equation}\label{dfun}
 \left.d(k,s,s')^{\pm1}\right|_{(k,s,s')\sim(\pm2\sIR,\sIR,\sIR)}
  ~=~  e^{
 -2\pi i(s^2+s'^2)-\frac{i\pi}{4} +if(\pi(s+s'\mp k))}\, ,
\end{equation}
where the function $f(x)$ is defined as\footnote{ The function $f(x)$
is related to the odd function $g(x)$ from Appendix A of \cite{Malong}
by $f(x) = \hf \pi x^2 + \frac{\pi}{12}- g(x)$.}
\begin{equation}
 f(x) ~=~ \frac1\pi\int_{-\infty}^xd\zeta
 \left(\frac{\zeta}{\tanh\zeta}+\zeta\right),
\label{Svf}\end{equation}
see Appendix A for details.  
\def\dWS{  \delta_{\rm WS}}

\subsection{Three-point amplitude}

In order to compute the four-point function we need the expression of
the three-point function in the Maldacena limit.  It gives the
amplitude of a long folded string emitting or absorbing a short open
string.  To the lowest order, the computation boils down to that of
three-point function of boundary operators $B_\beta\equiv
e^{\beta\phi}$ in Liouville theory on a disk.  The corresponding
structure constant has been worked out by \cite{Ponsot:2001ng} but the
general formula is quite complicated.  It actually simplifies in
Maldacena's limit when the conservation of energy,
$k_1+k_2+k_3=-\tilde Q/2$, is taken into account.  We evaluate this
structure constant as a common solution of the shift relations derived
in Appendix B,\footnote{A variant of these relations  has been previously derived  by V. Petkova \cite{ValyaP}.}
\begin{eqnarray}
\lefteqn{
 \vev{^{s_3}[U_{k_1-ib/2}^-]^{s_1}[U_{k_2}^-]^{s_2}[U_{k_3}^-]^{s_3}} 
+\vev{^{s_3}[U_{k_1}^-]^{s_1\pm i b/2}[U_{k_2-ib/2}^-]^{s_2}[U_{k_3}^-]^{s_3}} 
} \nn\\&&
 ~=~
 \frac{\nuu u ^Q\pi^2b^{\frac12}\cdot
       d(k_1,s_3,s_1\pm b/2)^{-1}\, d(k_2,s_1,s_2)^{-1}\, d(k_3,s_2,s_3)^{-1}}
      {\sin(2\pi ik_1b)\, \sin(2\pi ik_2b)}\, 
\label{3ptrec1}
\end{eqnarray}
and  
\begin{eqnarray}
\lefteqn{
 \vev{^{s_3} [U_{k_1+i/2b}^+]^{s_1}[U_{k_2}^+]^{s_2}[U_{k_3}^+]^{s_3}} 
+\vev{^{s_3}[U_{k_1}^+]^{s_1\pm i/2b}[U_{k_2+i/2b}^+]^{s_2}[U_{k_3}^+]^{s_3}} 
} \nn\\&&
 ~=~
 \frac{\nuu u ^Q\pi^2b^{-\frac12}\, d(k_1,s_3,s_1\pm i/2b)\, d(k_2,s_1,s_2)d(k_3,s_2,s_3)}
      {\sin(2\pi ik_1/b)\, \sin(2\pi ik_2/b)}\, .
\label{3ptrec2}
\end{eqnarray}

Let us first solve the relation (\ref{3ptrec1}) in Maldacena's limit
taking $\{k_1,k_2,k_3\}\sim\{+2\sIR ,0,-2\sIR \}$.  The term on the
right hand side scales as
\[
 {\rm r.h.s.} ~\sim~ e^{4\pi i\sIR \left(s_1'-s_2+k_2+ib^{-1}\right)}.
\]
One of the two terms on the l.h.s. has to scale in the same way, and the
other has to be subdominant or comparable.  By inspection one finds,
\[
\begin{array}{lclclcl}
 s_1=s_1'+\tfrac{ib}2 &\Longrightarrow& {\rm l.h.s.}1&\ll&{\rm
 l.h.s.}2&\sim&{\rm r.h.s.}, \\
 s_1=s_1'-\tfrac{ib}2 &\Longrightarrow&
 {\rm l.h.s.}1&\sim&{\rm l.h.s.}2&\sim&{\rm r.h.s.}.
\end{array}
\]
So (\ref{3ptrec1}) reduces to a two-term relation for
$s_1=s_1'+\tfrac{ib}2$ and is easily solved.  The solution, when
transformed into the amplitude of $U_k^+$, reads
\begin{equation}
 \left.\vev{^{s_3}[U^+_{k_1}]^{s_1}[U^+_{k_2}]^{s_2}[U^+_{k_3}]^{s_3}} 
 \right|_{(k_1,k_2,k_3)\sim(2\sIR ,0,-2\sIR )}
 ~=~
 \frac{4i\pi^2\nuu  ^Qb^{\frac12}}
      {e^{2\pi b(k_1+s_2)}-e^{2\pi b(-k_3+s_1)}}.
\label{B3pt1}
\end{equation}
By a little more work it can be shown that this solution satisfies all
nonequivalent recursion relations which follow from (\ref{3ptrec1}).
It is also easy to see that (\ref{B3pt1}) satisfies the homogeneous
version of the recursion relation (\ref{3ptrec2}), i.e. the equation
with the r.h.s. set to zero.

The second recursion relation (\ref{3ptrec2}) can be analyzed in the
same way, and one can find a solution which in terms of $U_k^-$ reads
\begin{equation}
 \left.\vev{^{s_3}[U^-_{k_1}]^{s_1}[U^-_{k_2}]^{s_2}[U^-_{k_3}]^{s_3}} 
 \right|_{(k_1,k_2,k_3)\sim(2\sIR ,0,-2\sIR )}
 ~=~
 \frac{-4i\pi^2\nuu ^Qb^{-\frac12}}
      {e^{\frac{2\pi}b(k_1+s_2)}-e^{\frac{2\pi}b(-k_3+s_1)}}.
\label{B3pt2}
\end{equation}
This solution is easily seen to satisfy the homogeneous version of
(\ref{3ptrec1}). The correct three-point amplitude in Maldacena's limit 
is thus given by the sum of the two expressions (\ref{B3pt1}), (\ref{B3pt2}).

\bigskip

 \noindent   {\bf The two-point amplitude.}

To make a precise comparison between the results of worldsheet
computations and Matrix Quantum Mechanics, we need a precise form of
the disk two-point amplitude.  It turns out slightly different from
the reflection coefficient (\ref{dkss}) for the operators $U_k^\pm$ by
a $k$-dependent function.  Computing the two-point amplitude from the
first principle is rather difficult; the CFT correlator is divergent
due to the zero-mode integrals of the fields $X$ and $\phi$, and it
has to be divided by the infinite volume of the residual global
conformal group that fixes the disk with two boundary insertions.  A
simple way to avoid these infinities is to differentiate with respect
to the boundary cosmological constant to make it a disk three-point
amplitude, where the additional boundary operator has the energy
$-\tilde Q/2$.

We recall that the three-point amplitude with $k_2=-\tilde Q/2$ can be
expressed in terms of the two-point function, see {\it e.g.} App.  D
of \cite{BGR}.  This relation can be extended for complex momenta by
solving the recursion relations (\ref{3ptrec1}), (\ref{3ptrec2}) for
$k_1= k= -k_3, k_2=-\tilde Q/2 $.  The result is
\begin{equation}
 \vev{^{s_3}[U_k^+]^{s_1}[U_{-k}^-]^{s_2}[{\rm c}B_b]^{s_3}}
 ~=~ -\frac{\pi\nuu^Q}{\sin(2\pi ik/b)}\cdot
      \frac{d(k,s_1,s_2)-d(k,s_1,s_3)}
           {\muB (s_2)-\muB (s_3)}.
\end{equation}
This is indeed a derivative with respect to $\muB $ when $s_2=s_3$.
We integrate it with respect to $\muB $ assuming that the naive
integration is allowed only when the operators satisfy Seiberg's bound
\cite{Seiberg:1990eb}
\[
 \epsilon\cdot{\rm Im}k > 0 ~~~~{\rm for}~~~U^\epsilon_k.
\]
The resulting two-point amplitude becomes non-analytic,
\begin{equation}
 \vev{^{s_2}[U_k^+]^{s_1}[U_{-k}^-]^{s_2}}
 ~=~ {\rm sgn}({\rm Im}k)\ \frac{\pi\nuu^Q d(k,s_1,s_2)}{\sin(2\pi ik/b)}.
\label{b2pt}
\end{equation}

\subsection{Boundary Ground Ring}

Similar recursion relations among higher point disk amplitudes can be
derived by making use of the boundary ground ring.  The ring is
generated by the operators $a_\pm$,
\begin{eqnarray}
 a_+ &=& -\nuu ^{-\frac{1}{2b}}
          \{{\rm bc} -\tfrac b2(\partial\phi+\partial X)\}
          e^{-\frac{1}{2b}(\phi-X)},\nn\\
 a_- &=& -\nuu  ^{-\frac b2}
         \{{\rm bc} -\tfrac{1}{2b}(\partial\phi-\partial X)\}
         e^{-\frac{b}{2}(\phi+X)}.
\end{eqnarray}
where ${\rm b, c}$ are the reparametrization ghost and antighost
fields.  Note that, since they are constructed from Liouville
degenerate operators, $a_\pm$ can only join two branes whose $s$
labels differ by a certain amount.  The operator products of $a_\pm$
with boundary tachyons satisfy the following formulae
\begin{equation}
-U_k^-a_- = a_-U_k^- =  U_{k-\frac{ib}2}^-, ~~~
-U_k^+a_+ = a_+U_k^+ =  U_{k+\frac{i}{2b}}^+.
\label{BOPE}
\end{equation}
The ring relation ($a_+a_-=a_-a_+=1$ at $b=1$) is realized on physical
open string operators in much the same way as for the bulk ground
ring, except that $a_\pm$ also shifts the label of the brane.  The
above formulae are simple and independent of the labels of branes,
whereas the coefficient of the OPE $a_\pm U^\mp_k$ becomes a little
complicated and can be obtained from (\ref{BOPE}) by reflection,
\[
 {}^s[a_-]^{s'}[U_k^+]^{s''} ~=~ {}^s[U_{k-\frac{ib}{2}}^+]^{s''}\cdot
 \frac{d(k,s',s'')}{d(k-\frac{ib}2,s,s'')}.
\]

By inserting an $a_\pm$ in a disk amplitude and using the fact that
$\partial a_\pm$ is BRST-exact, one can derive a shift relation among
disk amplitudes.  As an example, consider the difference of four-point
amplitudes
\[
 \vev{U^\pm_{k_0}(U^\pm_{k_1} a_\pm)U^\pm_{k_2} U^\pm_{k_3}}
-\vev{U^\pm_{k_0} U^\pm_{k_1}(a_\pm U^\pm_{k_2})U^\pm_{k_3}}.
\]
Since this can be written as an integral of an amplitude containing
$\partial a_\pm$, it vanishes by BRST invariance up to contributions
from the boundary of moduli space of disks with marked points.  As was
discussed originally in \cite{Bershadsky:1992ub}, see also \cite{BGR},
such boundary contributions are summarized by the higher operator
products
\begin{eqnarray}
 a_-U_{k_1}^+U_{k_2}^+ &=&
+\frac{b^{\frac12}\pi}{\sin(2\pi ibk_1)}
 U_{k_1+k_2-\frac i{2b}}^+, \nn\\
 U_{k_1}^+a_-U_{k_2}^+ &=&
+\frac{ b^{\frac12}\pi\sin(2\pi ib(k_1+k_2))}{\sin(2\pi ibk_1)\sin(2\pi ibk_2)}
 U_{k_1+k_2-\frac i{2b}}^+, \nn\\
 U_{k_1}^+U_{k_2}^+a_- &=&
-\frac{b^{\frac12}\pi}{\sin(2\pi ibk_2)}
 U_{k_1+k_2-\frac i{2b}}^+, \nn\\
 a_+U_{k_1}^-U_{k_2}^- &=&
-\frac{b^{-\frac12}\pi}{\sin(2\pi ik_1/b)}
 U_{k_1+k_2+\frac{ib}2}^-,\nn\\
 U_{k_1}^-a_+U_{k_2}^- &=&
-\frac{b^{-\frac12}\pi\sin(2\pi i(k_1+k_2)/b)}
      {\sin(2\pi ik_1/b)\sin(2\pi ik_2/b)}
 U_{k_1+k_2+\frac{ib}2}^-,\nn\\
 U_{k_1}^-U_{k_2}^-a_+ &=&
+\frac{b^{-\frac12}\pi}{\sin(2\pi ik_2/b)}
 U_{k_1+k_2+\frac{ib}2}^-.
\label{BCONT1}
\end{eqnarray}
Using them, the four-point amplitude can be shown to satisfy the
recursion relation,\footnote{ In our convention for disk amplitudes
$\vev{U_1\cdots U_n}$, the first the second, and the last operators
are unintegrated and the rest are integrated (with the factor of $\rm
c$ removed).  }
\begin{eqnarray}
\lefteqn{
 \vev{U^\pm_{k_0}U^\pm_{k_1}(a_\pm U^\pm_{k_2})U^\pm_{k_3}}
-\vev{U^\pm_{k_0}(U^\pm_{k_1}a_\pm) U^\pm_{k_2}U^\pm_{k_3}}
} \nn\\&=&
 \vev{(U^\pm_{k_0}U^\pm_{k_1}a_\pm) U^\pm_{k_2}U^\pm_{k_3} }
-\vev{ U^\pm_{k_0}(U^\pm_{k_1}a_\pm U^\pm_{k_2})U^\pm_{k_3}}
-\vev{ U^\pm_{k_0}U^\pm_{k_1}(a_\pm U^\pm_{k_2}U^\pm_{k_3})},
\label{rec4pt}
\end{eqnarray}
where the operator products in the parentheses are given by
(\ref{BOPE}) and (\ref{BCONT1}).

Although the formulae for the operator products were derived in
\cite{Bershadsky:1992ub, BGR} in the theory without Liouville
interaction, we assume they remain valid after it is turned on.  The
interaction will, however, make higher operator products $a_\pm U^n\to
U~ (n\ge 3)$ non-vanishing as well.  The determination of higher point
amplitudes along this path will therefore become more and more
difficult.

\subsection{Four-point amplitude (scattering of two long strings)}

We will evaluate  the amplitude describing the scattering of two long strings, 
\[
 \vev{^{s_3}[U^-_{k_0}]^{s_0}[U^+_{k_1}]^{s_1}[U^-_{k_2}]^{s_2}[U^+_{k_3}]^{s_{3}}} ;~~~~~
 (k_0,k_1,k_2,k_3)~\sim~(-2\sIR  ,2\sIR,-2\sIR,2\sIR),
\]
as the common solution of the two recursion relations (\ref{rec4pt}).
Both relations (\ref{rec4pt}) have inhomogeneous terms on the right
hand side.  As was the case with three-point amplitude, we can find
the solution by working with those inhomogeneous terms one by one and
then combining the results together in a manner consistent with the
symmetry.  Consequently, the four point amplitude will consist of a
number of terms.

\begin{figure}[htb]
\centerline{\includegraphics{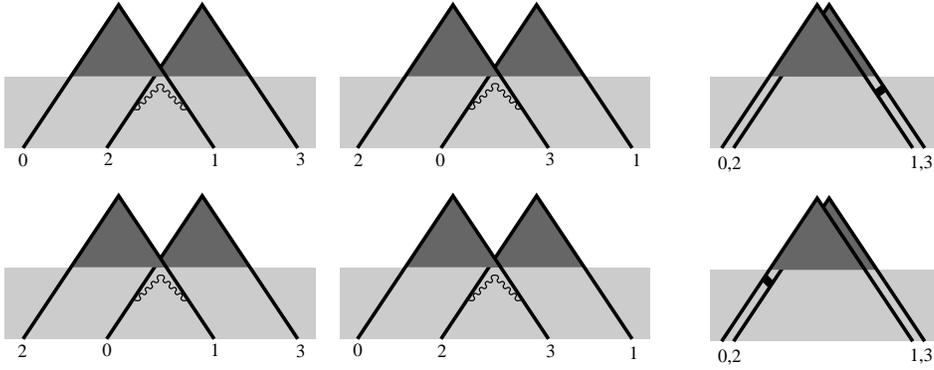}}
  \caption{ \small{Terms in disk four-point amplitudes.  The four
  figures on the left correspond to physically uninteresting processes
  and are discarded.  The two on the right describe the terms which
  are reproduced from MQM analysis.  } }
\label{fig:fourpt}
\vskip 15pt
\end{figure}

Some of the terms read,
\begin{eqnarray}
\lefteqn{
\vev{^{s_3}[U^-_{k_0}]^{s_0}[U^+_{k_1}]^{s_1}[U^-_{k_2}]^{s_2}[U^+_{k_3}]^{s_{3}}}}
\nn\\ &=& 16\pi^3\nuu ^Q \frac{d(k_3,s_2,s_3)}{d(k_0,s_3,s_0)}
\frac{d(k_0+k_1+\tfrac{\tilde Q}{2},s_3,s_1) \sin 2\pi
ib(k_0+k_1+\tfrac{\tilde Q}2)} {(e^{2\pi b(k_1+s_3)}-e^{2\pi
b(-k_0+s_1)}) (e^{\frac{2\pi}b(k_3+s_1)}-e^{\frac{2\pi}b(-k_2+s_3)})}
\nn\\&+& 16\pi^3\nuu ^Q \frac{d(k_3,s_2,s_3)}{d(k_2,s_1,s_2)}
\frac{d(k_1+k_2+\tfrac{\tilde Q}{2},s_0,s_2) \sin 2\pi
ib(k_1+k_2+\tfrac{\tilde Q}2)} {(e^{2\pi b(k_1+s_2)}-e^{2\pi
b(-k_2+s_0)}) (e^{\frac{2\pi}b(k_3+s_0)}-e^{\frac{2\pi}b(-k_0+s_2)})}
\nn\\&+& 16\pi^3\nuu ^Q \frac{d(k_1,s_0,s_1)}{d(k_2,s_1,s_2)}
\frac{d(k_2+k_3+\tfrac{\tilde Q}{2},s_1,s_3) \sin 2\pi
ib(k_2+k_3+\tfrac{\tilde Q}2)} {(e^{2\pi b(k_3+s_1)}-e^{2\pi
b(-k_2+s_3)}) (e^{\frac{2\pi}b(k_1+s_3)}-e^{\frac{2\pi}b(-k_0+s_1)})}
\nn\\&+& 16\pi^3\nuu ^Q \frac{d(k_1,s_0,s_1)}{d(k_0,s_3,s_0)}
\frac{d(k_3+k_0+\tfrac{\tilde Q}{2},s_2,s_0) \sin 2\pi
ib(k_3+k_0+\tfrac{\tilde Q}2)} {(e^{2\pi b(k_3+s_0)}-e^{2\pi
b(-k_0+s_2)}) (e^{\frac{2\pi}b(k_1+s_2)}-e^{\frac{2\pi}b(-k_2+s_0)})}
\nn\\ &+&\cdots.
\label{4ptsol1}
\end{eqnarray}
From their dependence on the reflection amplitude $d$, they seem to
describe the processes in which a short open string is exchanged
between the incoming leg of one long string and the outgoing leg of
the other, as described by the left four of Figure \ref{fig:fourpt}.
These terms will therefore be physically uninteresting and discarded.
Indeed, we will see these terms are not reproduced from MQM. Moreover,
in Maldacena's limit these terms are subdominant and infinitely
rapidly oscillating as compared to the terms which are reproduced from
the MQM.

The physically interesting terms, Figure \ref{fig:fourpt} right, are
obtained from the recursion relations with the number of inhomogeneous
terms reduced,
\begin{eqnarray}
&& ~\hskip-9mm
 \frac{d(k_1,s_0,s_1')}{d(k_1-\frac{ib}2,s_0,s_1)}
 \vev{^{s_3}[U^+_{k_0}]^{s_0}[U^+_{k_1-\frac{ib}2}]^{s_1}[U^+_{k_2}]^{s_2}[U^+_{k_3}]^{s_3}} 
 \nn\\&& ~\hskip-12mm
+\frac{d(k_2,s_1,s_2)}{d(k_2-\frac{ib}2,s_1',s_2)}
 \vev{^{s_3}[U^+_{k_0}]^{s_0}[ U^+_{k_1}]^{s_1'}[U^+_{k_2-\frac{ib}2}]^{s_2}[U^+_{k_0}]^{s_3}} 
\nn\\ &=&
- \frac{8\pi^3\, \nuu ^Q\, b \, e^{-2\pi bk_1}}
       {e^{2\pi b(k_3+s_1)}-e^{2\pi b(-k_2+s_3)}}
+ \frac{8\pi^3\, \nuu ^Q\, b \, e^{2\pi bk_2}}
       {e^{2\pi b(k_1+s_3)}-e^{2\pi b(-k_0+s_1)}}
\end{eqnarray}
and a similar equation involving $i/2b$ shifts.  We found that they
are solved for $b=1$ by
\begin{eqnarray}
\lefteqn{\vev{^{s_3}[U^-_{k_0}]^{s_0}[U^+_{k_1}]^{s_1}[U^-_{k_2}]^{s_2}[U^+_{k_3}]^{s_{3}}} 
}
\nn\\ &=&  16i\pi^3\nuu ^2
   e^{-2\pi k_2}\times
   \frac{d(k_1,s_0,s_1)d(k_3,s_2,s_3)
        -d(k_0,s_3,s_0)^{-1}d(k_2,s_1,s_2)^{-1} }
     {(e^{2\pi(k_1+s_2)}-e^{2\pi(-k_2+s_0)})
      (e^{2\pi(k_3+s_1)}-e^{2\pi(-k_2+s_3)})}
 \nn\\&=&
 4i\pi^3\nuu ^2  \, e^{-\pi( k_1+k_3)}\times 
  \frac{d(k_1,s_0,s_1)d(k_3,s_2,s_3)
        -d(k_0,s_3,s_0)^{-1}d(k_2,s_1,s_2)^{-1} }
     {\sinh \pi(k_1+k_2-s_0 +s_2) \sinh(k_2+k_3+s_1-s_3)
      }
.\nn\\ 
\label{4ptsol2}
\end{eqnarray}

A remarkable property of (\ref{4ptsol2}) is that it has a certain
symmetry under the exchange of $k_a$ and $s_a$.  To see this, let us
introduce four positive ``winding'' parameters $y_0, y_1, y_2,y _3$
such that
\eqn\ksigm{
k_0 = -(y _3+y _0),~~~
k_1 =   y _0+y _1 ,~~~
k_2 = -(y _1+y _2),~~~
k_3 =   y _2+y _3.
\la{ksigm}}
Then the conservation of momenta is satisfied automatically.  The
parameters $y_i$ are determined up to a common translation
\eqal\trany{ y_{0,2}\to y_{0,2} + a, \qquad y_{1,3}\to y_{1,3}-a \, .
}
By inserting them into (\ref{4ptsol2}) one finds that in Maldacena's
limit the non-trivial part of the amplitude depends only on the
differences $ \hat y_i =  y_i- s_i$:
     \begin{eqnarray}\la{fourp}
   \vev{^{s_3}[U^-_{k_0}]^{s_0}[U^+_{k_1}]^{s_1}
   [U^-_{k_2}]^{s_2}[U^+_{k_3}]^{s_{3}}}      
              &= &
  4\pi^3\, \nuu ^2\,  \prod_{j=0}^3 e^{- 2\pi i s_j^2 - \pi(s_j+y_j)}
    \nn\\
 & & \hskip -2.9cm \times\  \frac{ e^{if(-\pi\hat y_0+\pi\hat y_1)
            +if(-\pi\hat y_2- \pi\hat y_3)}
         -e^{-if(\pi \hat y_3-\pi\hat y_0)
            +if(-\pi\hat y_1-\pi\hat y_2)}}
        { \sinh \pi ( \hat y_2-\hat y_0)
       \  \sinh\pi(\hat y_1-\hat y_3)}
.
\end{eqnarray}
In other words, the four-point amplitude is almost symmetric under
$s_i-\sIR \leftrightarrow \sIR-y _i$.  The change of sign can be
understand as follows: increasing the energy makes the tip of the
folded string go further while increasing the boundary parameter has
the opposite effect.  The parameter $y$ is in some loose sense T-dual
to the  original time variable; such a duality transformation is
discussed for the AdS disk amplitudes in \cite{AM}.

 \section{Long folded strings in Matrix Quantum Mechanics}
\label{sec:mqm}

\subsection{Asymptotic states and chiral formalism of MQM}

The dual matrix description of the 2D string theory in the linear
dilaton background is given by a dimensional reduction of a 2D YM
theory to one dimension, known also as Matrix Quantum Mechanics
\cite{reviews}.  The theory involves one gauge field ${\bf A} =
\{A_{i}^j \} $ and one scalar field ${\bf X} =\{X_{i}^{j} \}$, both
hermitian $N\times N$ matrices.  It is formally defined by the action
\eqn\Maction{ \CS= \int dt \, \tr \left[ {\bf P}\, \nabla_{\bf A}{\bf
X} -
\hf ({\bf P}^2- {\bf X}^2) 
\right] ,
\label{Maction}}
where $\nabla_{\bf A }{\bf X} = \p_t {\bf X} -i [{\bf A}, {\bf X}]$ is
the covariant time derivative.  The action (\ref{Maction}) can be
considered as an effective action describing the states near
a local maximum of a confining potential for the scalar field.  
In this approximation a  generic potential can be replaced by 
inverse gaussian potential and a large cutoff parameter $\Lambda$.  
The number of colors $N$ should be tuned appropriately with the 
cutoff $\Lambda$ before taking the large $N$ limit.

The Hilbert space of MQM decomposes as a direct sum 
\eqn\HiSp{ \cH  = \cH _0 + \sum_{n=1}^\infty \cH _n \, ,
 \label{HiSp}
 } 
where $\cH _0 $ is the singlet sector and the sector $\cH _n $ is
obtained by adding $n$ Wilson lines in the adjoint representation.
The sector $\cH _n $ can be further decomposed into a direct sum of
irreducible representations of $U(N)$ whose Young tableaux contain $n$
boxes and $n$ `antiboxes'.  The Young tableaux for the allowed
representations wih $n=1,2$ are represented in Fig.  3.

In the singlet sector, $n=0$, the action (\ref{Maction}) describes a
system of $N$ non-relativistic free fermions in the upside-down
quadratic potential \cite{reviews}.  The ground state of the system is
characterized by the Fermi level $E_F=-\mu$.  The cutoff $\Lambda$
then gives the energy (with minus sign) of the $N$-th level below the
surface of the fermi sea.  In the large $N$ limit, the non-singlet
sectors can be described in terms of impurities in the fermi sea.  The
non-singlet excitations of MQM have been studied in \cite{GKv, BULKA,
KKK}.  Let $\CD$ be an allowed irreducible representation of $SU(N)$.
Then the radial part of the Hamiltonian contains a term with Calogero
type interaction between the eigenvalues,
{\begin{equation}
H^{ (\CD )}=-\frac{1}{2}\sum_{j=1}^N \left(
{\partial^2 \over \partial x _j^2}+x _j^2
\right)+\frac{1}{2}\sum_{j\neq k}^N \frac{\CD(E_{j}^{k})\CD(E_{k}^{j})}{
(x _j-x _k)^2}
\, .
\la{Hadj}
\end{equation}
where  $\CD(E_{j}^{k})$ is a realization of  the $N\times N$ matrix 
$(E_{j}^{k})_{l}^{m}=\delta_{j}^{m}\delta_{l}^{k}$.

In the adjoint representation, $n=1$, the wave function is an $N\times
N$ traceless matrix.  By an $SU(N)$ rotation it can be diagonalized as
$$\Psi_{i}^j (\X)= \delta_{i}^j\ \psi_i (\X),
$$
with $\sum_i \psi_i =0$.  Then the radial part of the Schr\"odinger
equation closes on the components $\psi_1,\dots, \psi_N$:
\eqn\intactp{
 H^{\rm adj} \psi_i  =  
- \sum_{j=1}^N {\hf} 
( \p_{i}^2  +x_i^2)\psi_i +
 \sum_{  j(\not= i)}^N { (\psi_{i} - \psi_j) \over (x_i - x_j)^2 } 
 \, .
\la{intactp}
 }
The solution of this equation in the large $N$ limit \ gives the
wavefunction of one quasiparticle.
 
Once we know the solution of the wave equation in the adjoint, which
was found in \cite{Fid}, we can try to explore the higher sectors.  To
the leading order in $1/\mu$, the sector $\cH _n$ describes $n$
non-interacting quasiparticles.  The statistics of the quasiparticles
is determined by choice of the irreducible representation.  Our
analysis of the Calogero type wave equation (see Appendix E) shows
that the Hamiltonian indeed decomposes into a sum of terms
(\ref{intactp}) associated with the $n$ quasiparticles, and a two-body
interaction Hamiltonian, which is of order $1/\mu$.

Instead of trying to solve the wave equation, in this paper we will
follow an alternative approach, the chiral quantization of MQM
\cite{AKK}, which proved to be very efficient in the singlet sector.
Since the potential is unbounded, any observable can be formulated in
terms of scattering amplitudes between asymptotic states that
characterize the system at the infinite past and in the infinite
future.  Usually the scattering matrix relating the incoming and the
outgoing asymptotic states is extracted from the asymptotics of the
solution of the Schr\"odinger equation.  In the case of quadratic
potential it happens that the $S$-matrix can be constructed directly,
without passing through the evaluation of the wave function.  This is
possible due to the important property of the MQM that the asymptotic
in- and out-states depend on the light cone variables
\eqn\YpYm{ \Xp = \frac{ {\bf X}+ {\bf P}}{ \sqrt{2}}, \ \ \ \ \quad
X_- = \frac{ {\bf X}- {\bf P}}{ \sqrt{2}}.  }
For example, the operators
\eqn\tachy{ T^+_{\kk}= \tr \, \Xp^{i\kk}, \qquad T^-_{\kk}= \tr \, \Xm^{-i\kk}.
} describe the left- and right-moving tachyons with energy $\kk$
\cite{Jevicki}.  The time evolution of the asymptotic states is
governed by the Hamiltonian
\eqn\ophac{ H=-\frac{1}{2}\tr ( \Xp\Xm + \Xm \Xp) \, , \label{opham}}
and the general solution of the corresponding Schr\"odinger equation
is
\eqn\GENSOL{ \Phi^{\pm} (\Xpm , t) = e^{\mp {1\over 2} N^2 t} \
\Phi^{\pm}(e^{\mp t}\Xpm ).
 \label{GENSOL}
 }
Thus any homogeneous function is an eigenstate of the Hamiltonian
(\ref{opham}).

\begin{figure}[tbp]
	\begin{center}
		\includegraphics[width=9cm]{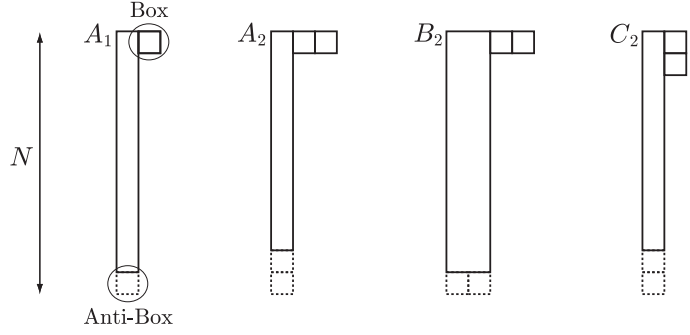}
	\end{center}
	\caption{ \small{The lowest representations allowed in MQM.  }	}
	\label{fig:young}		
\end{figure}

The outgoing and the incoming states are related by matrix Fourier
transformation
\eqal\Fouriertr{ \Phi^+(\Xp) &= \ \int d\Xm e^{- i\tr \,\Xp\Xm}\,
\Phi^-(\Xm)\, ,}
which represents the scattering operator in the chiral basis. 

The wave functions in the sector $\cH _n$ transform according to the
$n$-th power of the adjoint representation,
\eqn\Unrot{ \Phi ^\pm(\O \Xpm\O^\dagger)= {\rm Ad} (\Omega)^{\otimes
n} \, \Phi^\pm(\Xpm), \qquad \O\in SU(N).
\label{Unrot}
 }
For any $n$, ${\rm Ad} (\Omega)^{\otimes n} $ decomposes into a direct
sum of irreducible representations.  The projection to any given
irreducible representation is obtained by applying the corresponding
Young symmetrizer.  The scattering amplitude between the states $ \Phi
^-$ and $ \Phi ^+ $ is given by the inner product
\eqn\inner{ \( \Phi^+, \Phi^- \)= \int d \Xp d\Xm \ e^{ i \tr\, \Xp
\Xm} \  \tr^{_{\!\! (n)}} \,  \overline{ \Phi^+ ( \Xp)}  
\Phi ^- (\Xm)
\label{inner}\, ,
}
where $\tr^{_{\!\! (n)}} $ denotes the  trace in the $n$-th tensor product.

With this description of the non-singlet sectors we can represent any
state as a polynomial of the matrix fields $\Xp$ or $\Xm$ multiplying
a singlet wave function.  Below we will see that similarly to the 
closed strings, the long folded strings are represented in MQM 
 by  creation and annihilation operators made of the matrix elements of  $\Xp$ and $\Xm$.   Thus  the eigenfunction representing
a state with $n$ folded strings and $m$ tachyons is
\eqn\genst{[ \Phi^{\pm}( \kk_1,..., \kk_n; \kk'_1,...,\kk'_m )] _{j_1,
..., j_n}^{l_1,..., l_n} = \prod_{a=1}^n [ \Xpm^{\pm i \kk ^\pm_a}
]_{_{j_a}}^{^{l_a}} \ \prod_{b=1}^m \tr \Xpm^{\pm i \kk'_b} \ \ \Phi_0
^\pm (\Xpm)\, ,
\label{genst}
}
where $ \Phi_0 ^\pm $ is the ground state wave function.\footnote{Such
states form a complete, but not orthogonal set.  An orthonormal basis
of eigenstates is labeled by  the   irreducible representations of
$SU(N)$. 
The construction of such a basis was considered,  in the case of MQM with ``upside-up'' gaussian potential, in \cite{HM}.
  The eigenstates from the two sets are linearly related.  The
advantage of the first set of states is that the $N$ limit is easier
to construct, as well as the direct interpretation of these states in
terms of the worldsheet theory.  } 

 The scattering amplitude for such
states is given by the inner product of an incoming and an outgoing
state of the form (\ref{genst}) divided by the inner product of the
left and right ground states.  It is therefore useful to introduce for
each pair of functions $F^+(\Xm), F^-(\Xp)$ the expectation value

 \eqn\defsp{ \lmu F^+
 F^- \rmu:= { \( \Phi_0^+ F^+, F^-\Phi^- _0 \) \over \( \Phi^+_0 ,
 \Phi_0^-\) }.  \la{defsp} }

In this way all observables in the non-singlet sector of MQM can be
obtained as multi trace correlators of an effective two-matrix model
with non-confining potential $\tr \Xp\Xm$. 
 The observables in this ``non-compact''
matrix model can be evaluated by an appropriate regularization.  It
happens that some of the results obtained for the usual, ``compact''
matrix models, can be applied here.

The $U(N)$ symmetry allows to reduce the original $N^2$ degrees of
freedom to the $N$ eigenvalues $ x^\pm_1\dots x^\pm_N$ of the matrices
$\Xp$ or $\Xm$.  The evaluation of the scattering amplitude is
achieved in two steps \cite{Iadj}.  The first step consists in
integrating out the angular degrees of freedom in the matrix
integration measure
\eqn\measre{ d\Xpm = d \O_\pm \, dx^\pm_1\dots dx^\pm_N\,
\Delta^2(x_\pm), }
where $ \Delta(x_\pm)= \prod _{i<j} (x^\pm_i- x^\pm_j)$ is the
Vandermonde determinant.  The angular integral in the inner product
(\ref{inner}) is then of the form
\eqn\GIZ{ I_{n} (\Xp, \Xm) = \int \limits_{SU(N)} \!\!\!  d \O\
 {\rm Ad}(\O)^{\otimes n}  \ e^{i \Xp \O \Xm \O^{\dagger}}.
\label{GIZ}
} 
The second step is to take the large $N$ limit and express the
result in terms of the collective field, the phase space eigenvalue
density $\rho(x^+, x^-)$.  In the well studied singlet sector, $ n=0$,
the integral (\ref{GIZ}) is the Harish-Chandra-Itzykson-Zuber integral
\cite{HCIZ}.  In this simplest case the transition amplitudes
$$ S(\kk^+_1,..., \kk_{m}^+; \kk^-_1, ..., \kk^-_n)  
 =   \langle \mu \vert
\prod_{a=1}^m \tr \Xp^{i\kk^+_a} \ \prod_{b=1}^n\tr \Xm ^{ i\kk^-_b}\vert
\mu\rangle  \nn $$
then can be formulated in terms of the scalar product in the fermi sea
vacuum \cite{AKK}. 
In the general case the integral (\ref{GIZ}) was
evaluated by Shatashvili \cite{Shatash}. In the case $n=1$ a formula 
suitable for taking the large $N$ limit was guessed by Morozov 
\cite{moroz} and proved later by Eynard and collaborators 
\cite{BE, EynM}. More general integrals and and other gauge
 groups were studied in \cite{EF, FEFZ}.  Taking the  large $N$ 
 expansion of these exact results is a delicate task.  In this aspect,
 the paper \cite{EO} proved to be very useful for our problem.

 \subsection{The reflection amplitude in the  adjoint sector $(n=1)$}

The sector with $n=1$ contains only one non-trivial representation,
the adjoint.  One can think of this sector as the fermi sea in
presence of an impurity, or quasiparticle.  In terms of the string
theory, the adjoint sector describes incoming and outgoing asymptotic
states containing one folded string.  In absence of tachyons the inner
product in this sector gives the reflection amplitude of the
quasiparticle associated with the tip of the folded string
\cite{Iadj}.  In the compactified Euclidean theory this sector
describes states with one vortex and one anti-vortex.  The
eigenfunction describing a folded string with energy $E$ is of the
form
\eqn\wfgee{[ \Phi^{\pm} ( \kk ) ]_{j}^{l}= \left([ \Xpm^{\pm i \kk }
]_{_{j}}^{^{l}} -\frac{1}{N}\mbox{tr} (\Xpm^{\pm
i\kk})\delta_{_{j}}^{^{l}} \right) \Phi_0 ^\pm (\Xpm),
\label{wfn1}
}
where $ \Phi_0 ^\pm $ is the ground state wave function.  To the
leading order in the large $N$ limit one can replace $SU(N)$ by $U(N)$
and neglect the term subtracting the trace.  More general wave
functions,
$$
 [  \Xpm^{\pm i
\kk }  ]_{_{j}}^{{l}}  \  \tr \Xpm^{\pm i \kk_1}    \tr \Xpm^{\pm i \kk_2} \cdots
\tr \Xpm^{\pm i \kk_m} \ \Phi_0 ^\pm (\Xpm)\, ,
$$
describe a folded string in presence of $m$ tachyons.   We  will focus on
the states of the form (\ref{wfn1}).  For such
states the scattering matrix reduces to the reflection factor for one
adjoint particle, which is given by the normalized inner product
%
  \eqn\defRE{ R_1(\kk_+, \kk_-) =\lmu \tr (\Xp^{i \kk_+}
  \Xm^{i\kk_-})\rmu\, .  }

Evaluating the integral over the angles by the Morozov-Eynard formula, 
we obtain an expression depending only on the eigenvalues of the 
matrices $\Xp$ and $\Xm$. In the large $N$ limit the result can be expressed in terms of the joint eigenvalue density $\rho(\xp,\xm)$ for 
the ground state. The Morozov-Eynard formula  takes most simple
form \cite{EynM} when expressed in terms of the resolvents 
\eqn\defW{ \Wpm(\xipm) : = {1\over \xipm + {\bf X}_\pm}\, .  \la{defW}
}
The operators creating eigenstates with given energy (\ref{wfn1}) are
related to the operators (\ref{defW}) by the integral transformation
 \eqal\inbb{ {\bf X}^{-i\kk} = {i\over\pi} \sinh \pi \kk \int \limits
 _0^\infty { d \xi \over \xi + {\bf X}}\, \, \xi ^{-i \kk} \, .
\label{Mellinb}
}
The inverse transformation is
\eqal\inba{ {1\over \xi+ {\bf X}}=- {i\over 2 \xi} \int \limits _{C}
{d\kk\over\sinh\pi \kk} \ \xi^{i\kk} \, {\bf X}^{-i\kk}\, ,
\label{Mellina}}
where the integration contour $C$ is parallel to the real axes and
passing between the poles at $\kk=0$ and $\kk=i$ of the integrand.  It
is most natural to choose $C= \RR+\hf i$, which we will do in the
following.

We therefore first evaluate the normalized inner product for the
resolvents  (\ref{defW}),
\eqn\defRad{ G_1(\xip,\xim) := \lmu \tr [ \Wp(\xip)\Wm(\xim)]\rmu\, ,
\la{defRad} }
and then apply the integral transformation (\ref{Mellinb}) to obtain
the reflection amplitude in the $E$-space.  The expectation value 
(\ref{defRad}) is the basic single-trace mixed correlator  in the 
effective two-matrix model we mentioned before.
The result, obtained 
in \cite{Iadj}, is surprisingly simple: 
 \medskip
 \eqal\intph{  G_1(\xip,\xim)&= &
 e^{ - i S(\xip, \xim)}\, ,  \nn \\  & & \nn \\
 S(\xip.\xim)&=  &   \int {d\xp d \xm \over 2\pi} \ { \rho(\xp, \xm ) \over (
 \xip+\xp)( \xim +\xm)}
\, ,
\label{intph}
 }
where $\rho(\xp,\xm) $ is the semiclassical eigenvalue density of the
fermionic liquid.
%


This integral is logarithmically divergent and needs a regularization.
We introduce a cutoff $\tL\gg\mu$ as the depth of the Fermi sea
explored by the average.\footnote{This means that we consider
non-singlet excitations that transform according to a smaller group
$SU(\tilde N)\subset U(N) $ with $\tilde N\sim \tL\log\tL$.} We assume
that $\tL\ll \Lambda$, so that we still can use the spectral density
for the upside-down harmonic oscillator.  The part of the Fermi sea
that corresponds to the interval of energies $-\mu<E<-\tL$ is
described by the density function
\eqn\statro{ \rho(\xp,\xm) =\theta(\xp\xm-\mu)\, \theta(\tL-\xp\xm) .
}
The result of the integration with this density depends only on the
product $\xip\xim$:
\eqal\Sexp{  S(\xip, \xim)= f\(T-t\) - f
\(-t\) \, ,
\label{Sexp}}
where we denoted
\eqn\defM{ T = {1\over 2} \log{\tL \over \mu}, \qquad t = {1\over 2}
\log{\xip\xim \over \mu }
 \label{defM}
 }
and the function $f$ is the same as in (\ref{Svf}).  Applying the
integral transformation (\ref{Mellinb}) to both arguments of ${G_1}
(\xip,\xim) $, we get
\eqn\phasee{ 
R_1({ \kk}_+ , { \kk}_- )  =- \frac{\sinh(\pi{ \kk}_+)\sinh(\pi{ \kk} _-) 
}{\pi^2} \int_0^{\infty} d\xip d\xim (\xip)^{i{\kk} _+}
( \xim)^{i{ \kk} _-} \ G_1 (\xip,\xim) .  }
 At this point we change the variables as $ \xi^\pm = \sqrt{\mu} \,
 e^{t\pm \tau}\, .  $ The integral over $\tau$ produces a delta
 function imposing the energy conservation,
\eqal\phass{ R_1({\kk}_+ ,{ \kk}_- )= \CR_1 ({ \kk}_+ ) \ \delta({ \kk}
_+-{\kk} _-) \, .
   \label{phass}
}
The reflection factor for one quasiparticle is given by the remaining
integral in $t$:
\eqn\phasE{ \CR_1 ({ \kk}) =-{2\over \pi} \sinh^2\!  \pi\kk \, \,
\mu^{-i{\kk } +1}\, \int_{-\infty}^{\infty} d t \, e^{2 t (-i{ \kk}
+1) - i f( T-t) + i f(- t)}\, .
\label{phasE}
}
In the limit $T\to\infty$ we can use the approximation $f(T-t) =
(T-t)^2/\pi$ and then write the exponent, using (\ref{simfone}), as
$2t - i f(t) - 2i(\kk-T/\pi) t + i\pi/6$.  The integral is evaluated
using the last equation (\ref{ftxf}).  The final result for the
reflection factor is
   \eqn\phasEf{  
    \CR_1 ({ \kk}) = 2 \Lambda\,   e^ {-i \a}     \sinh^2\!\pi\kk  \, 
  \mu^{-i \kk}  \, 
   e^{-i T^2/\pi}  e^{-2\kk\pi + i f(T-\pi \kk)}
\, 
\label{phasEf}\, .
} 
%

We see that the reflection factor depends on the shifted energy $\eh =
E-T/\pi$, where the constant $T= {1\over 2} \log \tL /\mu$ is in fact
the logarithmic energy gap between the singlet and the adjoint sector
discovered in \cite{GKv}.  We therefore subtract, as in \cite{Malong},
this constant from the energy and introduce the shifted energy
variables
\eqn\defeh{\label{defeh} \e = \kk - {1\over 2\pi} \ln \tL, \qquad
\hat\e = \e +{1\over 2\pi} \ln{\mu } = {\kk} -T/\pi \, ,}
where $\eh$ is assumed finite.  Then we can approximate $\sinh \pi
\kk\approx {1\over 2} e^{\pi \kk}$ and the scattering phase takes the
form \eqal\phasEff{ \CR_1 ({ \kk}) &=& \ \hf \Lambda \, e^{ -i\a} \,
\mu^{-i{\kk} }\ e^{-i T^2/\pi} e^{ i f(-\pi\eh)}\cr && \cr &=& - \hf
\tL \, e^{-i\pi/4} e^{- i \log^2{\tL }/4\pi} \times e^{ i \pi \e ^2 -
i f(\pi\eh)} \, .
\label{phasef}
}

Let us compare this expression with the two-point function in the
worldsheet theory with $s=s'=M$ and $k\sim M$, 
\eqal\cmprs{ \< U^+_k
U^-_{-k}\>= {\pi \nu^2\over \sinh 2\pi |k|} d(|k|) = \pi^2 \nu^2 \, 2
e^{-2\pi k} e^{-i\pi/4} e^{- 4\pi i M^2} e^{i f(2\pi M-\pi k)}\, .  
}
Remarkably, the two expressions coincide (up to a constant phase,
which can be absorbed in the normalization of the wave functions) upon
the identification
 \eqal\identf{ k=E,\qquad s=s'= {T/
2\pi},\qquad \nu^2 = \pi\mu\, . 
\la{identfone}
 }
Therefore the cutoff $\tL$, the depth of the fermi sea felt by the
collective excitation, is related to the boundary cosmological
constant in the worlsheet theory:
\eqal\tLmB{
\tL = \muB^2\, .
\la{tLmB}
}
The remaining factor can be absorbed into the normalization of the
boundary operators $U^\pm_{\pm k}$.

Note that the reflection factor (\ref{phasef}) is given, up to a
complex conjugation and a numerical factor, by the same function as the mixed correlator
(\ref{defRad} ):
    \eqal\DUAone{ \CR_1(E)= \tL^{1-iE} \ \overline{G_{1}(
     \xip, \xim )} \qquad
     \xip\xim = \tL\,  e^{ -\pi E} = \mu\, e^{ -\pi \eh}\, .
     \la{DUAone} }

 \subsection{The reflection amplitude in the sector  $n=2$}

Now we will evaluate the reflection amplitudes for the states
(\ref{genst}) with $n=2$, in the leading and in the subleading order.
The scattering matrix is not diagonal for such states.  It gets
diagonalized in the basis of the irreducible representations with
$n=2$, which we describe below.

In the sector $\cH _2$ there are four irreducible representations
(denoted as $A_2, \bar A_2, B_2$ and $C_2$ in \cite{BULKA}).  Their
Young tableaux are shown in Fig.  3.  For general $n$, the
representations $B_n$ and $C_n$ are defined by tensors with $n$ upper
and $n$ lower indices, respectively totally symmetric and totally
antisymmetric under permutations of the upper and lower indices,
associated with boxes and antiboxes.  The representations $A_n$ are
totally symmetric in boxes and totally antisymmetric in antiboxes, and
similarly for $\bar A_n$.  The zero weight states in the sector with
$n=2$ are of the form
 $$
     \Psi_{ik}^{jl}
     = P_{ik}^{jl} \ \psi_{ik}\, ,
 $$
where $P_{ik}^{jl}$ is a standard tensor associated with the
corresponding Young symmetrizer.  In the four irreducible
representations, $A_2, \bar A_2, B_2$ and $C_2$, it is given
respectively by $ \d_{[i}^{\{j}\d_{k]}^{l\}} ,
\d_{\{i}^{[j}\d_{k\}}^{l]} ,  \d_{\{i}^{\{j}\d_{k\}}^{l\}}$ and
 $\d_{[i}^{[j}\d_{k]}^{l]}$, where $[\cdot , \cdot ]$ denotes
antisymmetrization and $\{\cdot , \cdot \}$ denotes symmetrization.
In order to extract the irreducible part, one needs further to impose
the tracelessness condition for any pair of upper and lower indices,
but this can be skipped in the two leading orders in the large $N$
limit.
      
Now let us return to the states (\ref{genst}) with $n=2$.  As in the
case $n=1$, it is advantageous first to evaluate the inner product of
the wave functions in the coordinate space
\eqn\deftPhid{ \tilde \Phi _{_{j_1,j_2; k_1,k_2}} ^{\pm}(\xipm_1, \xipm_2) =
[\Wpm(\xipm_1)]_{_{j_1}}^{^{k_1} } [\Wpm(\xipm_2)]_{_{j_2 k_2}}\, 
\Phi^\pm_0\, .  \la{deftPhid} }
The inner product is expressed in terms of mixed two-trace correlator
$$
G_{1,1}(\xip_1,  \xim_1; \xip_2,\xim_2)
=  \lmu \tr[W_+(\xip
_1) W_-(\xim_1)]\ \tr[W_+(\xip _2) W_-(\xim_2)] \rmu 
$$
and the mixed one-trace correlator
$$
G_{2}(\xip_1,  \xim_1, \xip_2,\xim_2)
=
 \lmu \tr [W_+(\xip _1) W_-(\xim_1) W_+(\xip _2)
 W_-(\xim_2)] \rmu \, .
 $$
The projections to the four irreps in the sector $n=2$ are obtained by
(anti)symmetrization:
\eqal\RRABS{ 
\langle \mu  |\tilde \Phi^{+}  \   \tilde \Phi^{-} 
| \mu\rangle_{A_2}
&=& 
G_{1,1}(\xip_1,  \xim_1; \xip_2,\xim_2)   -  \{
\xim_1\leftrightarrow \xim _2\}
\ =
\ \langle \mu  |\tilde \Phi^{+}  \   \tilde \Phi^{-} 
| \mu\rangle_ {\bar A_2}
\nn\\
\langle \mu  |\tilde \Phi^{+}  \   \tilde \Phi^{-} 
| \mu\rangle_{B_2}
  &=& 
  G_{1,1}(\xip_1,  \xim_1; \xip_2,\xim_2)  
  +G_{2}(\xip_1,  \xim_1, \xip_2,\xim_2) + \{ \xim_1\leftrightarrow \xim _2\}\, ,
 \nn \\
    \langle \mu  |\tilde \Phi^{+}  \   \tilde \Phi^{-} 
| \mu\rangle_{C_2}
    &=& 
  G_{1,1}(\xip_1,  \xim_1; \xip_2,\xim_2)  
  -G_{2}(\xip_1,  \xim_1, \xip_2,\xim_2) + \{ \xim_1\leftrightarrow \xim _2\}\, 
    .
    }
Each term on the r.h.s. can be expanded in $1/\mu$.  To the leading
order $ \mu^2$ only the first term contributes, where it factorizes to
$$ G_{1,1}(\xip_1, \xim_1; \xip_2,\xim_2) = G_1(\xip_1, \xim_1)\,
G_1(\xip_2,\xim_2)\, + O(\mu^0) .
$$
The integral transformation (\ref{Mellinb}) gives \eqal\Rlead{
R_2^{(0)} (E_1^+, E_2^+, E_1^-,E_2^-) = \d(E_1^+-E_1^-)\,
\d(E_2^+-E_2^-) \CR_1 (E_1^+)\, \CR_1 (E_2^+)\, .  \la{Rlead} } To this
order the inner product are given by the (anti)symmetrized product of
two expectation values (\ref{defRad}), associated with each trace.
This is the approximation of dilute gas of quasiparticles.

The subleading term is given by
$G_2(\xi_{1}^{+},\xi_{2}^{+},\xi_{1}^{-},\xi_{2}^{-}) $.  Here we will
focus on this amplitude and leave the next order $\mu^0$, which is
given by the connected correlator of the product of two traces, for
future work.  The mixed one-trace correlators $G_n$ in the effective
two-matrix model can be expressed through the lowest one-trace
correlator $G_1$ by the general formula derived in \cite{EO}.  In the
case $n=2$ it states
\eqal\EquR{G_2(\xi_{1}^{+},\xi_{2}^{+},\xi_{1}^{-},\xi_{2}^{-})
&=&i\, {G_1 (\xi_{1}^{+},\xi_{1}^{-})G_1(\xi_{2}^{+},\xi_{2}^{-})
-G_1(\xi_{1}^{+},\xi_{2}^{-})G_1(\xi_{2}^{+},\xi_{1}^{-})
\over(\xi_{1}^{+}-\xi_{2}^{+})(\xi_{1}^{-}-\xi_{2}^{-}) } \, .
\la{EquR} }
A simpler derivation of this formula, due to L. Cantini, can be found
in Orantin's thesis \cite{Orantin}.  The advantage of this derivation,
which we give in Appendix C, is that it can be applied also for our
non-compact effective two-matrix model.

The next step is to perform the integral transformation
(\ref{Mellinb}) to each of the arguments.  The calculation, this time
non-trivial, is presented in Appendix B. The result contains a
$\d$-function for the energy conservation, so we define the subleading
$n=2$ reflection amplitude $\CR_2$ by
\eqal\RtwoE{ R_2^{(1)}(\kk_1^+,\kk_2^+, \kk_1^-,\kk_2^-) =
\d(\kk_1^++\kk_2^+- \kk_1^--\kk_2^-)\ \CR_2(\kk_1^+,\kk_2^+;
\kk_1^-,\kk_2^-) \, .  \la{RtwoE} }
For energies $\kk _j^\pm = \e_j^\pm + {1\over 2\pi}\ln\tL $, with
$\e_j^\pm$ finite,
\eqal\CRtwo{ \la{Rtwo} \CR_2(\kk _1^+,\kk _2^+, \kk _1^-,\kk _2^-) &=&
- {2   \over\tL }  \ \e^{\pi
(\kk _1^++\kk _2^+)}\ { \CR_1 (\kk _1^+)\CR_1 (\kk _2^+)-\CR_1 (\kk _1^-) \CR_1 (\kk _2^-) \over
\sinh\pi(\kk ^+_1-\kk ^-_1) \sinh\pi(\kk ^+_1-\kk ^-_2) } \cr &&\cr&&\cr
&=&- \frac{2}{\mu} \ e^{ \pi(\eh^+_1+\eh^+_2)} \ \frac{
{\CR_1 }({\kk }_{1}^{+}) {\CR_1 }({\kk }_{2}^{+})- {\CR_1 }({\kk }_{1}^{-})
{\CR_1 }({\kk }_{2}^{-})
}{\sinh{\pi(\eh_{1}^{+}-\eh_{1}^{-})}\sinh{\pi(\eh_{1}^{+}-\eh_{2}^{-})}}
\, . 
}
The $n=2$ reflection amplitude, given in the first two orders by
(\ref{Rlead}) and (\ref{RtwoE}), obviously satisfies the unitarity
condition
\eqn\unitc{ \CR_1 (\kk _1^+) \CR_1 (\kk _2^+) \, \overline{ \CR _2( \kk _1^-
,\kk _2^-, \kk _1^+, \kk _2^+)} = \CR_2( \kk _1^+ ,\kk _2^+; \kk _1^-,
\kk _2^-)\, \overline{ \CR_1 (\kk _1^-) \CR_1 (\kk _2^-)} \la{unitc} .}

Let us compare the subleading reflection amplitude (\ref{Rtwo}) with
the 4-point disk amplitude in the Maldacena limit (\ref{4ptsol2}),
evaluated in the worldsheet theory.   If we identify 
 \eqal\identftwo{&& k_0 =- E^-_1, 
 \quad k_1 = 
  E^+_1,\quad  k_2 =- E^-_2, 
  \quad k_3 = E^ +_2
 ,\cr &&
 \quad s_0=s_1=s_2=s_3 = {T/
2\pi},\quad \nu^2 = \pi\mu\, ,
\la{identftwo}
 }
 then the two amplitudes are indeed equal to each other, up
to a factor of 2.  This factor can be absorbed in the normalization of
the functional measure in the worldsheet calculation.  After fixing
the normalization of the boundary operators and the functional
measure, there are no more ambiguities left.

     If we express the  energies in terms of the shifted  winding parameters      
    $  y_i  $  defined by (\ref{ksigm}),
 \eqal\ehy{
 E _1^-=   y_3+  y_0,\quad
  E _1^+=   y_0+  y_1,\quad
  E _2^-=   y_1+  y_2,\quad
    E _2^+=   y_2+  y_3\, ,
 \la{ehy}   }
then we observe that  the reflection amplitude in the $E$-space has again   the same functional form as
 in the $\xi$-space,
  \eqal\GtwoEt{
  \mathcal{R}_{2}(E_{1}^{+},E_{1}^{-},E_{2}^{+},E_{2}^{-}) =-2i
  \tL^{2-i(E_{1}^{+}+E_{2}^{+})} \
  \overline{G_{2}( \xip_1,\xim_1,\xip_2,\xim_2  )}\,
   \la{DUAtwo} }
  with
\eqal\xiy{&& \xim_1= {\tL ^{1/2}} \, e^{-{1\over 2} \pi   y_0},\quad 
\xip_1= {\tL ^{1/2}}\, e^{-{1\over 2}  \pi   y_1},
\cr && \cr &&
\xim_2=  {\tL ^{1/2}}  \, e^{-{1\over 2}  \pi   y_2},\quad 
\xip_2= {\tL ^{1/2}} \,e^{-{1\over 2} \pi   y_3}
\,.
\la{xiy}}
%
%

%

\subsection{ The reflection amplitude for $n>2$}

Using the loop equations for the two-matrix model one can evaluate all
mixed one-trace correlators
$$
{G}_n(\xip_1,\xim_1,\dots,\xip_n,\xim_n) := \langle \mu \vert \tr
\left[ W_+(\xip_1) W_-(\xim_1)\dots W_+(\xip_k) W_-(\xim_n)\right]
\vert \mu\rangle
.
$$
These correlators satisfy the recurrence equations found in \cite{EO},
see Appendix C. The unique solution of the recurrence equations is
given by the ``Bethe Ansatz like'' formula of \cite{EO} as a sum of
products of $G_1$ with rational coefficients:
\eqal\bethelike{ {G}_n(\xip_1,\xim_1,\ \dots,\xip_ n,\xim_ n) =
\sum_{\s \in \bar S_n} C_\s(\xip_1,\xim_1,\ \dots,\xip_ n,\xim_ n) \
G_1(\xip_1,\xim_{\s_1})\dots G_1(\xip_n,\xim_{\s_n}) \, .
\la{bethelike} } 
The sum goes in the set $\bar S_n$ of planar
permutations of $n$ elements.  A planar permutation $\s$ can be
defined as follows.  Consider a disk with $2n$ marked points on the
boundary, labeled by $\xim_1, \xip_1,\dots , \xim_n, \xip_n$.  Drow a
set of arcs (oriented lines) connecting the points $\xip_i$ with
$\xim_{\s_i}$, $i=1,\dots, n$.  The permutation $\s\in S_n $ is planar
if the arcs can be drown without intersections, as shown in Figure
\ref{fig:Rainbow}.  We call such a configuration of arcs ``rainbow
diagram''.

It is shown in \cite{EO} that the coefficient $C_\s$ is equal to a
product of weights associated with the $n+1$ windows on the rainbow
diagram.  These weights are determined from the following
recurrence relation, which stems from the loop equations
(\ref{loopeqW}).  Denote by $C_j(\xip_1, \xim_1, \dots, \xip_j,
\xim_j)$ the weight of the window with $2j$ points $\xip_1,\xim_1,
\dots, \xip_j, \xim_j$ along its boundary, labelled following the
orientation of the lines.  This function is invariant under cyclic
permutations of the $2j$ points.  Then the weight $C_n$ is expressed
through $C_1, \dots, C_{n-1}$ as
\eqal\recC{ C_n(\xip_1,\xim_1,\dots, \xip_n,\xim_n) = i\, \sum_{j=1}^n
{C_j(\xip_1,\xim_1,\dots, \xip_j,\xim_j)\
C_{n-j}(\xip_{j+1},\xim_{j+1},\dots, \xip_n,\xim_n) \over
(\xip_n-\xip_1)(\xim_n-\xim_j)}\, , }
where by definition
$
C_1(\xip,\xim) = 1$  \cite{EO}.

 \begin{figure}
\centerline{\includegraphics[width=65mm]{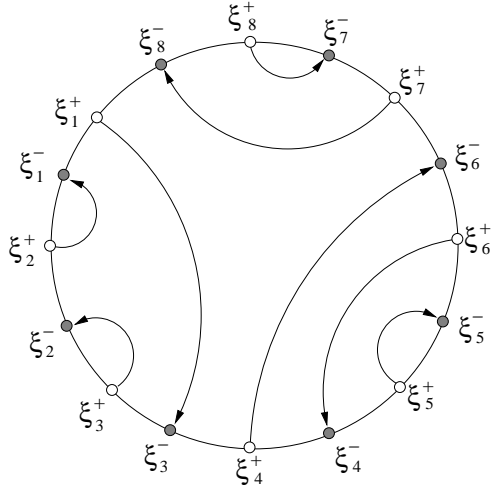}
 }
  \caption{  \small{  A planar set of arcs on the disk with $2j$ points 
  along the boundary, labeled by $\xip_1,\xim_1, \dots, \xip_j, \xim_j$,
  and its dual tree graph.
  On the figure $j=8$. The  arcs split the disk into $j+1$ windows.
  The non-trivial weights of the windows in the figure are  $ 
  C_2(\xip_7, \xim_8, \xip_8, \xim_7)$,
  $C_3(\xip_1,\xim_3, \xip_4, \xim_6, \xip_7, \xim_8)$,
 $ C_3(\xip_1,\xim_3, \xip_3,\xim_2,\xip_2, \xim_1),$
 $  C_2(\xip_4, \xim_6,\xip_6,\xim_4)$, $C_2(\xip_6,\xim_4, \xip_5,\xim_5).
  $
  } }
\label{fig:Rainbow}
\end{figure}

In order to evaluate the amplitude in the $E$ space, we have to
perform the integral transformation (\ref{Mellinb}) of the r.h.s. of
(\ref{bethelike}).  We believe, although we are not able to supply the proof now,
that the relations (\ref{DUAone}) and (\ref{DUAtwo}) hold in fact for
any $n$.  If  we introduce 
the winding parameters $ y_1, \dots, y_{2n-1}, y_{2n}\equiv y_0$,
determined up to a common translation
$ y_{2i}\to y_{2i} + a, \  y_{2i-1}\to y_{2i-1}-a$,
and express the energies as
\eqal\nyeh{
E^+_i =  y_{2i-2} +   y_{2i-1}\,,\qquad E^-_i = y_{2i-1} + y_{2i}\, ,
\la{nyeh}
}
we conjecture  that  for any $n$
    \eqal\RnE{
  \CR_n(E^+_1,E^-_1,\dots, E^+_n, E^-_n)
  = \tL^{n- i(E^+_1+\dots E^+_n)}\
  \overline{G_n( \xip_1,\xim _1 , \dots,  \xip_n,\xim_n)\, ,
  }
  }
where 
\eqal\xiyn{
\xip_i= \tL^{1/2} \, e^{ -{1\over 2}   {\pi}  y_{2i-2}},\qquad
\xim_i = \tL^{1/2} \,  e^{-{1\over 2}   {\pi}   y_{2i-1}}
 \qquad (i=1, \dots, n)\, .}
 The energies $E_i$ and the winding parameters $y_i$ are logarithmically divergent with  the cutoff  $\tL$.   In terms of the finite  shifted energies  $\eh^\pm_i$
 the above relations reads
 \eqal\nyeh{&&
\eh ^+_i =  \hat y_{2i-2} +   \hat y_{2i-1}\,,
\qquad \eh ^-_i = \hat y_{2i-1} + \hat y_{2i}\, , \cr
&&
\xip_i= \mu^{1/2} \, e^{ -{1\over 2}   {\pi}  \hat y_{2i-2}},\qquad
\xim_i = \mu^{1/2} \,  e^{-{1\over 2}   {\pi}   \hat y_{2i-1}}
 \qquad (i=1, \dots, n)\, .}
\la{nyeht}
}
%

From the perspective of the worldsheet theory, each term in the
solution (\ref{bethelike}) is the contribution of a particular
scattering process involving $n$ long strings sharing a common
worldsheet.  The arcs in the rainbow diagram on Figure
\ref{fig:Rainbow} correspond to the tips of the $n$ long strings.  The
factorization of the amplitude means that the interaction occurs only
in the far past ( $t\sim - \log\muB$) or in the future ( $t\sim
\log\muB$).  For $-\log\muB\ll t\ll \log\muB$ the $n$ long strings
evolve independently, contributing a product of reflection factors
$\CR_1 $.  Furthermore, the recurrence relation (\ref{fig:fourpt}) for
the coefficients $C_j$ means that each such coefficient can be written
as a sum of prodicts of $C_2$.  Therefore the interaction of the long
strings can be decomposed into exchanges of short strings as those
illustrated in Figure \ref{fig:fourpt}.

\section{Discussion}

In this paper we studied the simplest scattering processes of long
folded strings in the two-dimensional string theory.  We used the
description of the long strings in terms of a FZZT brane with
$\muB\gg\sqrt{\mu}$, as suggested in \cite{Malong}.  We evaluated the
four-point disk amplitude in Maldacena's limit of large momenta and
large boundary parameters.  We observed that the amplitude depends
only on the chiral combinations $\hat y_i = y_i-s_i$, where $s_a$ are
the Liouville boundary parameters and $y_a$ the ``winding parameters''
, associated with the four segments of the boundary ($a=1,2,3,4$).
Furthermore we observed a symmetry with respect to the Fourier
transformation in the time direction.  Our result suggests that, to 
the leading order,  the
long strings interact {\it via} exchange of short open strings
that happens in the asymptotically free region $\phi\ll -\log\mu$.

On the side of MQM the scattering amplitudes are expressed in terms of
the mixed trace correlators in an effective non-compact two-matrix
model with interaction $i\tr\Xp\Xm$.  However the worldsheet
interpretation of the correlation functions is not the same as in the
matrix model description of rational $c<1$ string theories.  In our
case the spectral parameter $\xipm$ is related by Fourier
transformation to the energy of the boundary tachyon, while the
boundary condition for the Liouville field is determined by a
regularization parameter $\tL$, the depth of the fermi sea.  This is
because the boundary is associated to the gauge field ${\bf A}$, while
the fields $\Xpm$ create the asymptotic open string states.  Since we
are modelling a Lorentzian theory, the worldsheet description does not
have a statistical interpretation as a sum over surfaces with positive
weights.

We evaluated the subleading amplitude for any number $n$ of long
strings using the results for the single-trace mixed correlator in
 the two-matrix model \cite{EO}.  
 We obtained the result in the coordinate space,
while the worldsheet calculation gives it in the space of energies.
We performed explicitly the Fourier transformation for the cases $n=1$
and $n=2$  where we  reproduced  the result of the worldsheet theory up to factors that can
be absorbed into the normalization of the boundary tachyons and the
integration measure.

We  found that the  Fourier transformed amplitude takes essentially the 
same form as that in the coordinate space,  after being expressed in
terms of the $2n$ dual  coordinates $y_i$ associated with the boundaries  of the disk.   We were able to establish this symmetry
 only for the cases $n=1$ and $n=2$, but we believe that  it is a 
 general property of the disk $n$-point amplitude in Maldacena's limit.

Eventually we are interested  taking the limit with a large  number  of FZZT branes, which would help us, as suggested in \cite{Malong},  to find a matrix model description of the Lorentzian black hole.  The results reported in this paper  suggest that  the effective two-matrix model that stems from the chiral quantization of MQM is the right tool to achieve this limit.   In this paper we studied the interactions of long folded strings due to exchange of short open strings, which are described by the single-trace mixed correlators in the effective two-matrix model.   We did not consider the interactions  due to the  exchange of  closed strings.  Such interactions are  described by multi-trace mixed correlators in the effective two-matrix model.

%
  
\bigskip \leftline{\bf Acknowledgments}

\noindent We thank S. Alexandrov and N. Orantin for valuable
discussions.  This work has been partially supported by the European
Union through ENRAGE network (contract MRTN-CT-2004-005616), ANR
programs GIMP (contract ANR-05-BLAN-0029-01).  Part of this work was
done during the ``Integrability, Gauge Fields and Strings'' focused
research group at the Banff International Research Station.

\appendix

\section{ Properties of the functions $\bS(z)$ and $f(x)$}

We give here a short summary of the properties of the double sine
function $\bS(z)$ used in \cite{FZZb} and the function $f(x)$
introduced (for $b=1$) in \cite{Malong}.  The double sine is related
to the double gamma function introduced by Barnes \cite{Ba1}.  The
function $e_b(x) = e^{i f(\pi x)}$ is also known as non-compact
quantum dilogarithm \cite{FK}.

\subsection{The double sine function $\bS(z)$
 }

\begin{itemize}
 \item Integral representation: \cite{FZZb}
\eqn\lnS{ \log {\bS}( Q/2-ix )\,=\, {i\over 2} \int\limits_{0}^\infty
\(\frac{\sin 2t x} {\sinh bt \, \sinh {b^{-1}t } }-{2x\over t}\)
\,\frac{dt}{t}\, , \qquad Q= b+1/b \, .  }
\item Functional relations: \eqn\funcS{ \frac{\bS(z+b)}{\bS(z)}=
{2\sin {\textstyle\pi b z} }\,, \quad \frac{\bS(z+1/b)}{\bS(z)}=
{2\sin\frac{\textstyle\pi z}{b}}\, .  \la{funcS} }
%
 \item Poles and zeroes:
$$
\mbox{poles at}\ \ z =n_1 b+n_2 /b, \; \; n_1,n_2\leq 0 ,\quad
\mbox{zeroes at}\ \ z=n_1 b+n_2 /b, \; \; n_1,\ n_2\geq 1\, .
$$

 
\item  Asymptotics  at infinity (for ${\rm Re}b>0$): 
\begin{equation}
  \bS(x) ~\sim~ \exp \( {\mp i\pi}/{2}\left\{ \left(x-\hf Q\right)^2-
  {\textstyle\frac{1}{12}(b^2+b^{-2})} \right\}\) ~~~~ ~~~~({\rm
  Im}x\to \pm\infty)\, .
\end{equation}

\end{itemize}

\def\CS{{\cal S}}

\subsection{The function $f(x)$}

 \begin{itemize}
\item Definition:
%
\eqn\deff{ e^{i f(x) }= {e^{i {x^2/2\pi } + i \pi (b^2 + b^{-2})/ 24}
\over \bS ( Q/2-ix/\pi )} \, .
\label{deff}
}
%
\item Integral representation: 
\eqn\intrf{ f(x)= -{i\over 4} \int_{\RR+i0} {dt\over t}\,
{e^{-2itx/\pi}\over\sinh tb \, \sinh t/b} \, .  \la{intrf} }

\item Functional equations: 
\eqal\symf{ f(x)+f(-x)&=& {x^2\over\pi} +{ \pi (b^2+b^{-2})\over 12}
\, ,\cr &&\cr e^{i f(x+ i \pi b)}& =& {e^{i f(x)}\over 1+ e^{i\pi b^2}
e^{2b x}} \, , \cr &&\cr e^{i f(x+ i \pi /b)} &=& {e^{i f(x)}\over 1+
e^{i\pi /b^2} e^{2x/b}} \, .  \la{symf} }

\item Asymptotics at infinity (for $b>0$):
\begin{eqnarray}\label{as1}
 f(x) \; \to \;\left\{
\begin{array}{ll}
0, & \Re x\to-\infty\\
  x^2/\pi +\pi   (b^2+b^{-2})/12  \ , & \Re x\to +\infty \, .
\end{array}\right.
\end{eqnarray}


\item
Fourier transform \cite{ FKV,KLS}
 \begin{eqnarray}\label{ftxf}
 \int _{\RR } {dt\over\pi} 
  \, e^{ -   Q t} \ e^{if(-  t)} \, e^{ 2  itx/\pi}
 &=& e^{-i  \a}\; e^{  Q x} \ e^{-i f( x)}\;
 \cr
 &&\cr
 \int _{\RR } {dt\over\pi}  \, e^{ -   Q t} \ e^{-if(-  t)} \, e^{ 2 itx/\pi }
 &=& e^{i  \a}\; e^{ -  Q x} \ e^{i f(- x)}\;
  \cr&&\cr
 \int _{\RR } {dt\over\pi} \, e^{    Q t} \ e^{if(  t)} \, e^{ 2  itx/\pi }
 &=& e^{-i  \a}\; e^{ - Q x} \ e^{-i f(- x)}\;
\cr &&\cr
 \int _{\RR } {dt\over\pi}  \, e^{   Q t} \ e^{-if(  t)} \, e^{ 2  izt/\pi}
 &=& e^{i  \a}\; e^{  Q x} \ e^{i f( x)}\;
\end{eqnarray}
where $\a= \pi{1+Q^2\over 12} - \pi {Q^2\over 4}$.

\end{itemize}

\subsection{The  limit  $b\to 1$
}

\begin{itemize}

\item Integral representation:

Evaluating the derivative 
\begin{equation}
 \frac{d}{dx}\log\bS(1-ix)|_{b=1}
  ~=~ i\int_0^\infty dt\left[
  \frac{\cos 2tx}{\sinh^2t}-\frac{1}{t^2} \right]
  ~=~ \frac{-i\pi x}{\tanh\pi x}.
\end{equation}
we find another integral representation of the dounle sine  for $b=1$:
\begin{equation}
 i\pi\log\bS(1-ix)|_{b=1}
  ~=~ \pi\int_0^x dx'\frac{\pi x'}{\tanh\pi x'}
  ~=~ \int_0^{\pi x} d\zeta\frac{\zeta}{\tanh\zeta}.
\end{equation}
 Then (\ref{deff}) reproduces the
  function $f(x)$ as defined in \cite{Iadj}:
\begin{equation}
 f(x) ~=~ \frac1\pi\int_{-\infty}^xd\zeta
 \left(\frac{\zeta}{\tanh\zeta}+\zeta\right).
 \end{equation}

\item
 Functional equations:
\eqal\symfone{\la{simfone} f(x)+f(-x)&=& {x^2\over\pi} +{ \pi \over 6}
\, ,} \eqal\foireh{ \qquad e^{i f(x- i \pi )} = (1 - e^{2 x}) \, e^{i
f(x)} \, , \la{symff} }


  \bigskip
\item  Boundary reflection coefficient $d(k,s,s')|_{b=1}$:
\begin{eqnarray}
 d(k,s,s') &\stackrel{b=1}=& \bS(1+2ik)
 \bS(1-i(k+s+s'))\bS(1-i(k+s-s')) \nn\\ &&\hskip14mm \times\,
 \bS(1-i(k-s+s'))\bS(1-i(k-s-s')).
\end{eqnarray}
\begin{eqnarray}
 \left.d(k,s,s')^{\pm1}\right|_{(k,s,s')\sim(\pm2M,M,M)}
  &=& e^{
 -2\pi i(s^2+s'^2)-\frac{i\pi}{4} +if(\pi s+\pi s'\mp\pi k)}.
\end{eqnarray}

\end{itemize}

\section{Equation for the boundary Liouville three point function}

 Here we derive a shift equation for the three-point function of
 boundary operators $B_\beta\equiv e^{\beta\phi}$ in Liouville theory
 on a disk.  We denote the relevant structure constant by
 $C^{\beta_1,\beta_2,\beta_3}_{~s_1,s_2,s_3}$:
\begin{eqnarray*}
\vev{B_{\beta_1}(x_1)B_{\beta_2}(x_2)B_{\beta_3}(x_3)}_{s_1,s_2,s_3}
&=&
C^{\beta_1,\beta_2,\beta_3}_{~s_1,s_2,s_3}
\prod_{\{ijk\}}|x_i-x_j|^{h_k-h_i-h_j}.
\end{eqnarray*}
Our notation is such that $B_{\beta_1}$ joins the branes $s_3$ and
$s_1$, $B_{\beta_2}$ joins $s_1$ and $s_2$ and so on.  The shift
equation follows from the bootstrap constraints of four-point
functions containing one boundary degenerate operator $B_{-b/2}$ or
$B_{-1/2b}$.  Note that the two branes joined by $B_{-b/2}$ have to
satisfy \cite{FZZb}
\[
 s-s' = \pm ib/2~~~~{\rm or}~~~~
 s+s' = \pm ib/2,
\]
and similarly for those joined by $B_{-1/2b}$.

To recall where the constraints arise from, let us consider a
four-point function,
\[
 \vev{B_{\beta_1}(x_1)B_{-b/2}(x)B_{\beta_2}(x_2)B_{\beta_3}(x_3)}
 _{s'_1,s_1,s_2,s_3}.
\]
Using the analytic solutions of Virasoro Ward identity with the
knowledge of the operator product expansions involving $B_{-b/2}$, one
can derive linear relations among $C^{\beta_1\mp
b/2,\beta_2,\beta_3}_{s_1,s_2,s_3}$ and $C^{\beta_1,\beta_2\mp
b/2,\beta_3}_{s'_1,s_2,s_3}$.  One of them reads, after replacing
$\beta_i$ by $\frac Q2-ik_i$,
\begin{eqnarray*}
\lefteqn{
-  \frac{\Gamma(1-2ibk_1)}
        {\Gamma(\alpha-2ibk_1)}\, 
   C^{\frac Q2-ik_1-\frac b2,\frac Q2-ik_2,\frac Q2-ik_3}
    _{s_1,s_2,s_3}
+  \frac{\Gamma(1+2ibk_2-\alpha)}{\Gamma(2ibk_2)}
  \,  C^{\frac Q2-ik_1,\frac Q2-ik_2-\frac b2,\frac Q2-ik_3}
    _{s'_1,s_2,s_3}
} \nn\\&&
 ~=~ \frac{\Gamma(1+2ibk_1)\Gamma(1+2ibk_2-\alpha)}
          {\Gamma(1+2ibk_1+2ibk_2-\alpha)\Gamma(\alpha)}
\,    \frac{d_L(\frac Q2-ik_1,s_3,s_1')}
        {d_L(\frac Q2-ik_1+\frac b2,s_3,s_1)}
   \, C^{\frac Q2-ik_1+\frac b2,\frac Q2-ik_2,\frac Q2-ik_3}
    _{s_1,s_2,s_3}.
\end{eqnarray*}
Here $\alpha\equiv\frac12+ib(k_1+k_2+k_3)$, and $d_L$ is defined in
(\ref{dkss}).  See \cite{Ponsot:2001ng} and \cite{Kostov:2003uh} for
more detail.  
A similar relation is obtained from the four-point function containing
$B_{-1/2b}$.  This kind of shift relations among correlators is often
powerful enough to determine the structure constants in Liouville
theory.

We translate the above relations on Liouville correlators
into a shift relation among the disk amplitudes of three $U^-_k$ or
three $U^-_k$ in two-dimensional string theory.  The conservation of
energy requires $k_1+k_2+k_3=-\tilde Q/2$ or $\alpha=0$ in the above.
We will suppress the corresponding delta function when writing down
the formulae for amplitudes.  The divergence of $\Gamma(\alpha)$ in
the right hand side is canceled by the bulk divergence
\cite{Ponsot:2001ng}  of
$C^{\beta_1,\beta_2,\beta_3}_{s_1,s_2,s_3}$ at
$\beta_1+\beta_2+\beta_3=2Q$:
\begin{eqnarray*}
 C^{\beta_1,\beta_2,\beta_3}_{s_1,s_2,s_3} &\sim&
 \frac{d_L(\beta_1,s_3,s_1)d_L(\beta_2,s_1,s_2)d_L(\beta_3,s_2,s_3)}
      {2Q-\beta_1-\beta_2-\beta_3}.
\end{eqnarray*}
So we obtain (\ref{3ptrec1}) and (\ref{3ptrec2}).

\section{Derivation of the  correlator $G_2$  (\ref{EquR}).}
\def\nablam{\nabla_{\! _{-}}}
\def\nablap{\nabla_{\! _+}}
\def\nablapm{\nabla_{\! _{\pm}}}
\def\Wpa{W_{\! _{+, 1}} }
\def\Wpb{W_{\! _{+, 2}} }
\def\Wma{W_{\! _{-, 1}} }
\def\Wmb{W_{\! _{-, 2}} }
\def\Wpmab{W_{\! _{\pm, a}} }
\def\Wpmbb{W_{\! _{\pm, a}} }

 To evaluate the correlator $G_2$ we use the identity, following from
 the translation invariance of the matrix measure,
\eqal\idmes{
\int d\Xp d\Xm  
 \tr   ( \nablap \Wp^{\ 2}  \Wm^{\ 2} \Wp^{\ 1} \Wm^{\ 1} )
  \,   \Phi_0^+   \Phi_0^- \, \  e^{i\tr{X_{+}X_{-}}} =0  .
\la{Wor}
}
Here $\nablapm= \p _{\Xpm } $ is the operator of matrix derivative,
$\Phi_0^+(\Xp)$ and $\Phi_0^-(\Xm)$ are the wave functions of the left
and right ground states in the singlet sector and $ \Wpm^{\ a} \,
(a=1,2) $ denote resolvents (\ref{defW}) with spectral parameters
$\xipm_a$:
 $$
 \Wpm^{\ a}\equiv \Wpm (\xipm_a) = {1\over \xipm_a+\Xpm}\, .
 $$
We commute the operator of derivative to the right using that it acts
on the resolvents as $[\nablapm]_{j}^k \cdot [\Wpm^{\ a}]_{j'}^{k'} =-
[\Wpm^{\ a}] _{j'}^k \, [\Wpm^{\ a}]_{j}^{k'}$.  As a result we obtain
the identity
  \eqal\Wordone{ && \lmu -\tr \Wp^{\ 2}\, \tr (\Wp^{\ 2} \Wm^{\ 2}
  \Wp^{\ 1} \Wm^{\ 1} ) - \tr (\Wp^{\ 2} \Wm^{\ 2} \Wp^{\ 1} )\, \tr(
  \Wp^{\ 1} \Wm^{\ 1} ) \rmu\cr && + \lmu \tr (\Wp^{\ 2} \Wm^{\ 2}
  \Wp^{\ 1} \Wm^{\ 1} ( i \Xm + \nablap \ln \Phi_0^+ \rmu=0.
  \la{Wordone} }
 In a similar way,  replacing the trace in  (\ref{Wor}) with 
 $\tr  ( \nablap \Wm^{\ 2}  \Wp^{\ 1} \Wm^{\ 1} \Wp^{\ 2} )$, we obtain another 
 identity,
 \eqal\Wordtwo{ && \lmu - \tr (\Wm^{\ 2} \Wp^{\ 1} )\, \tr ( \Wp^{\ 1}
 \Wm^{\ 1} \Wp^{\ 2} ) -\tr (\Wm^{\ 2} \Wp^{\ 1} \Wm^{\ 1} \Wp^{\ 2} )
 \, \tr \Wp^{\ 2} ) \rmu\cr && + \lmu \tr (\Wm^{\ 2} \Wp^{\ 1} \Wm^{\
 1} \Wp^{\ 2} ( i \Xm + \nablap \ln \Phi_0^+ \rmu=0.  \la{Wordtwo} }
 Now we subtract (\ref{Wordtwo}) from (\ref{Wordone}) and use that
 $[\Wp^{\ 2}, \nablap \ln \Phi_0^+]=0$.  As a result we have an
 identity that does not involve the ground state wave function:
  \eqal\Wordt{ && \lmu \tr ( \Wp^{\ 1} \Wm^{\ 2} )\, \tr ( \Wp^{\ 2}
  \Wp^{\ 1} \Wm^{\ 1} ) - \tr ( \Wp^{\ 1} \Wp^{\ 2} \Wm^{\ 2} )\, \tr(
  \Wp^{\ 1} \Wm^{\ 1} ) \rmu\cr && + i \lmu \tr (\Wp^{\ 2} \Wm^{\ 2}
  \Wp^{\ 1} \Wm^{\ 1} \Xm ) - \tr ( \Xm\Wm^{\ 2} \Wp^{\ 1} \Wm^{\ 1}
  \Wp^{\ 2}) \rmu=0.  \la{Wordt} } Applying the identities
$$ \Wpm ^{\ 1} \Wpm^{\ 2} = -\, \frac{ \Wpm^{\ 1} - \Wpm^{\ 2}
}{\xipm_1 - \xipm_2}, \qquad \Xpm \Wpm^{\ a} =1-\xipm \Wpm^{\ a} \,
$$
 we write (\ref{Wordt}) as
  \eqal\Wordq{ && \langle \mu |\frac{ \tr ( \Wp^{\ 1} \Wm^{\ 2} )\, \tr
  (\Wp^{\ 2} \Wm^{\ 1} ) - \tr (\Wp^{\ 1} \Wm^{\ 1} ) \tr (\Wp^{\ 2}
  \Wm^{\ 2} ) }{ \xip_1 -\xip_2} |\mu \rangle 
  \cr &&\cr && -i \langle \mu | \frac{ \tr (
  \Wp^{\ 1} \Wm^{\ 2} - \Wp^{\ 2} \Wm^{\ 2} - \Wp^{\ 1} \Wm^{\ 1} +
  \Wp^{\ 2} \Wm^{\ 1} )}{ \xip_1 -\xip_2}  |\mu \rangle\cr
   &&\cr && - i \lmu \tr   (\Wp^{\ 1} \Wm^{\ 1} 
   \Wp^{\ 2} \Wm^{\ 2}) (\xip_1 -\xip_2)  \rmu \,=0\, .  \la{Wordq} }
In the leading order in the $1/\mu$ expansion we can use the
factorization of the normalized expectation values, $\lmu \tr A\,\tr B
\rmu= \lmu \tr A \rmu\lmu \tr B \rmu$.  Setting
$$G_{2}(\xi_{1}^{+},\xi_{1}^{-},\xi_{2}^{+},\xi_{2}^{-})
=\lmu \tr{\Wp^{\ 1}\Wm^{\ 1}\Wp^{\ 2}\Wm^{\ 2}} \rmu,
$$
$$
G_{1}(\xi^{+},\xi^{-})=\lmu \tr(\Wp\Wm)\rmu-i,$$
 we retrieve   (\ref{EquR}):
$$ 
G_{2}(\xi_{1}^{+},\xi_{1}^{-},\xi_{2}^{+},\xi_{2}^{-})= i \,
\frac{G_{1}(\xi_{1}^{+},\xi_{1}^{-})G_{1}(\xi_{2}^{+},
\xi_{2}^{-})-G_{1}(\xi_{1}^{+},\xi_{2}^{-})G_{1}(\xi_{2}^{+},
\xi_{1}^{-})}{(\xi_{2}^{+}-\xi_{1}^{+})(\xi_{2}^{-}-\xi_{1}^{-})}
\, .
$$
Proceeding in the same way  one obtains a recursive formula for the 
one-trace  correlator $G_n$ \cite{EO} 
\eqal\loopeqW{ {G}_n(\xip_1,\xim_1, \dots,\xip_ n,\xim_ n) &&=
i\sum_{j=1}^{k-1} \frac{ {G}_j(\xip_ 1, \xim_1, \dots,\xip_j,\xim_ j)
{G}_{k-j}(\xip_ {j+1} , \xim_{j+1} ,\dots, \xip_k, \xim_ n) }{ (\xim_
n-\xim_ j)( \xip_k -\xip_1) } \cr & &\ -i \sum_{j=1}^{n-1} {
{G}_j(\xip_ 1, \xim_1, \dots,\xip_j,\xim_ k) {G}_{n-j}(\xip_ {j+1} ,
\xim_{j+1} ,\dots, \xip_k, \xim_ j) \over (\xim_ k-\xim_ j)( \xip_n
-\xip_1) }\, . \nonumber \\
\la{loopeqW} }
%

\section{Evaluation of the integral for $R_2(\kk _1^+,\kk _2^+,
\kk _1^-,\kk _2^-)$}
 
\def\Em{ \kk ^{-}} \def\Ep{ \kk ^{+}} \def\Epm{\kk ^{\pm}}
\def\ve{\varepsilon}

In order to evaluate the disk scattering amplitude in the space of
energies, we need to perform the integral transformation
(\ref{Mellinb}) with respect to all four variables of the amplitude
(\ref{EquR}):
\eqal\Meltr{ R_2(\Ep_1,\Ep_2, \Em_1,\Em_2) = \frac{1}{\pi^{4}}
\sinh \pi \Ep _{1} \ \sinh \pi \Ep_{2}\ \sinh \pi \Em_{1} \ \sinh \pi
\Em _{2}\cr \times \int _0^\infty d\xip_1 d\xip_2 d\xim_1 d\xim_2 \ \,
( \xip_1)^{-iE _{1}^{+}} ( \xip_2)^{-iE _{2}^{+}} ( \xim_1)^{-iE
_{1}^{-}} ( \xim_2)^{-iE _{2}^{-}}\
G_2(\xi_{1}^{+},\xi_{2}^{+},\xi_{1}^{-},\xi_{2}^{-}) \, .  \la{Meltr}
} 
We will use the fact that the integral transform of the two factors
$G_1$ in the expression (\ref{EquR}) for $R_2$ is already known.  We
express the r.h.s. of (\ref{Meltr}) in terms of 
\eqal\hRCR{ R_1 (\Ep,\Em) = \d(\Ep-\Em)\, \CR_1 (\Ep)  \la{hRCR}}
by applying the inverse transformation (\ref{Mellina}):
\eqal\invM{  G_1 (\xi^+,\xi^-) &=& -{1\over 4 \xi^+\xi^-} \int\limits
_{ \RR+ i/2}{d\Ep \over\sinh\pi \Ep} \ \xi_+^{i\Ep} {d\Em
\over\sinh\pi \Em } \ \xi_-^{i\Em} \CR_1(\Ep,\Em)
.\la{InvM} }
As a result, the original integral takes the form
\eqal\Meld{\la{Meld}  R_2 &=& \int _{\RR+i/2}  \prod_{\pm}
[d p^\pm_1 d p^\pm_2 \ \ K ( \Epm_1, \Epm_2 ; p^\pm_1, p^\pm_2) ] \
\CR_1  (p_1^+,p_1^-) \, \CR_1  (p_2^+,p_2^-) \, \, , \nn \\
  } 
where the kernel $K(\kk_1, \kk_2; y_1, y_2) $ is given by the integral
\eqal\KernK{\la{KernK} 
K ( \kk_1, \kk_2 ; p _1, p _2) &=&  \frac{A}{2
\pi^2 } \int_0^\infty {d \xi_1 d\xi_2\over \xi_1\, \xi_2}\ \ {
\det\limits_{jk} \xi_j ^{i (p_k - \kk_j ) } \over \x_1- \x_2 }\ \, .  }
The factor 
\eqal\defOm{ A =\prod_{ j= 1,2} { \sinh\pi \kk _j \over
\sinh\pi p_j }
 \label{defOm}
 } 
can be put equal to 1, as we will see later.  To evaluate the kernel,
change the variables as $ \xi_1 = e^{\tau+t}, \ \ \xi_2 = e^{ \tau-t} $,
\eqal\Kpm{\hskip -0.5cm A^{-1}\, K ( \kk_1, \kk_2 ; p _1, p _2) &= &
{i \over \pi^2} \, \int_{-\infty}^\infty d\tau dt\, e^{ i\tau (p_1+p_2
- \kk _1 - \kk _2 ) } e^{it ( \kk _2 - \kk _1) }\ {\sin (p_2 -p _1)t
\over e^{\tau }\sinh t } \cr &=& {2 \over \pi} i\, \d( \kk _1 + \kk _2
- p_1-p_2 +i) \ \int _{-\infty}^\infty d t \ { \cos(\kk_1-\kk_2)t\
\sin (p _1 -p _2)t \over\sinh t}\cr &=& 2\, i\, \d( \kk_1 + \kk_2 -
p_1-p_2 +i) \ { \sinh\pi(p_1-p _2) \over \cosh\pi(\kk_1-\kk _2) +
\cosh\pi (p_1-p_2)} \, .  \nn \\
\la{Kpm} }
After substituting (\ref{Kpm}) and (\ref{hRCR}) in the
integral (\ref{Meld}), three of the integrations are compensated by
$\d$-functions, the remaining $\d$-function imposes the conservation
of the total energy:
\eqal\RharR{  R_2 (\Ep_1,\Ep_2, \Em_1,\Em_2)= \d(\Ep_1+\Ep_2 -
\Em_1 - \Em_2)\, \CR_2 .  }
\def\Dpm{\Delta_\pm} 
It is convenient to introduce the independent variables $\kk, \Dpm$ and
$\e'$ by
\eqal\chvar{ \Epm_1= \kk + { \Dpm }, \ \ \ \Epm_2 = \kk - { \Dpm }, \ \ \
\ p_1= \kk+ \e' + \hf i ,\ \ p_2= \kk- \e' + \hf i\, , }
$\e'$ being the integration variable.  It is clear that in the
Maldacena limit
\eqn\Ehe{ \kk = {1\over 2 \pi} \ln \Lambda + \e \la{Ehe} }
with $\e, \Dpm$ finite, the factor (\ref{defOm}) can be replaced by 1.
Then we can write the remaining integral as
\eqal\IntF{\la{IntF} \CR_2 =4i\,  \int_{-\infty}^\infty d\e'\ {
\frac{\sinh^{2}{2\pi \e'}\ \ {\CR_1 }( \kk+\e' + \hf i) {\CR_1 }(\kk- \e' + \hf
i)}{ (\cosh{2\pi\Delta _{+}}+\cosh{2\pi
\e'})(\cosh{2\pi{\Delta}_{-}}+\cosh{2\pi \e'})}} .}

We are going to evaluate the integral (\ref{IntF}) as a contour
integral.  According to (\ref{phasef}) the integrand behaves at $\e'
\to\pm\infty$ as $\exp ( - i \pi \e'^2)$.  This leads to the choice of
a 8-like contour shown in Fig.  3.  We will show later that the
integral integration along the imaginary axis is zero, so that the
integral (\ref{IntF}) is given by the sum of the residues trapped
inside the 8-shaped contour.  The integrand contains two kinds of
poles.  First, there are the (simple) poles of the kernel at
$$\e'=\pm{\Delta}_{-}+(n+\hf )i, \ \pm{\Delta}_{+}+(n+\hf )i, \qquad n\in\ZZ .
$$

\begin{figure}
\centerline{\includegraphics[width=55mm]{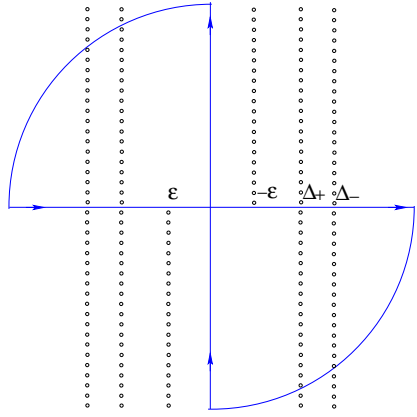}}
  \caption{  \small{ The 8-like contour in the $\e'$-plane and the
pattern of the poles. } }
\label{fig:Contur}
\end{figure}

\noindent Second, there are the poles of the factors ${\CR_1 }( E \pm \e'
+ \hf i)$ at
$$\e'=\mp \eh \pm (n+\hf)i, \qquad n\in\ZZ_{+}\, , $$
 where by $\eh$ we denoted the shifted energy $\eh = \e + {1\over 2\pi
 } \ln\mu = \kk- {1\over 2\pi } \ln(\Lambda /\mu)$.
The order of these poles grow linearly with $n$.

We will assume that $\eh<0$.  Then the second kind of poles remain
outside the integration contour and the only contribution will come
from the poles of the kernel.  Let us sum up the contributions of the
poles along the lines $\Re \e =\pm \Delta_-$.  Assume for definiteness
that ${\Delta}_{-}>0$.  Then we have to sum up the resudues at
\eqn\polesDp{ \e'=\pm {\Delta}_{-} \mp (n+\hf)i, \quad n\in\ZZ_+.
\la{polesDp} }
The residues of the  kernel are the same for all poles:
\eqn\ResK{ {\rm Res} K=i\,  
\frac{\sinh{2\pi\Delta_{-}}}{2\pi(\cosh{2\pi\Delta_{+}}-\cosh{2\pi\Delta_{-}})}
\, .  } 
The factor ${\CR_1 }{\CR_1 }$ in the integrand is evaluated at the $n$-th
pole using the shifting property
\eqn\shiftR{ {\CR_1 }(\kk+ni)= \mu^{-n}(1-e^{-2\pi\eh})^{-n}\, {\CR_1 }(\kk) = (
\mu -e^{-2\pi\e })^{-n}\, {\CR_1 }(\kk) \, .  }
The sum over the poles yields a factor 
\eqal\Dplus{
\sum_{n\in\mathbb{Z}_{+}}{\CR_1 (\kk+\Delta_{-}-ni)\CR_1 (\kk-\Delta_{-}+(n+1)i)}
\cr
=\frac{\CR_1 (\kk+\Delta_{-})\CR_1 (\kk-\Delta_{-})}{\mu(1-e^{-2\pi(\hat{\e}
-\Delta_{-})})}\sum_{n=0}^\infty
{\left(\frac{1-e^{-2\pi(\hat{\e}+\Delta_{-})}}{1-
e^{-2\pi(\hat{\e}-\Delta_{-})}}\right)^{n}} \cr
 = - e^{2\pi\e }\ {\CR_1 (E+\Delta_{-})\CR_1 (E-\Delta_{-})\over 2
 \sinh 2\pi \Delta_-}\, .  } 
Taking into account the contribution of both series of poles
(\ref{polesDp}) we get
\eqal\CRtwo{ \CR_{2}|_{{\rm poles}\ {\Delta}_{-}}= -{2 }\, 
e^{2\pi\e} \ \frac{ \CR_1  ( \kk+ {\Delta}_{-} ) \CR_1  ( \kk-
{\Delta}_{-} ) }{\cosh{2\pi {\Delta}_{+}}-\cosh{2\pi{\Delta}_{-}}} \,
.  }
Similarly we evaluate the contribution of the poles with $\Re \e=\pm
\Delta_+$.  Returning to the original variables $\Epm_j$ we write the
final result as
\eqal\CRfinal{ {\CR}_{2}=
-\frac{2}{\Lambda } \ e^{\pi(\Ep_1+\Ep_2) } \ \frac{
{\CR_1 }(\Ep_{1}) {\CR_1 }(\Ep_{2})- {\CR_1 }(\Em_{1})
{\CR_1 }(\Em_{2})
}{\sinh{\pi(\Ep_{1}-\Em_{1})}\sinh{\pi(\Ep_{1}-\Em_{2})}}
\, . }
 In terms of the renormalized energies  (\ref{defeh}),
 $\eh^\pm_j= \Epm_j - {1\over 2\pi}\ln(\Lambda/\mu)$  the result reads
  \eqal\CRfinalh{ {\CR}_{2} 
  = - \frac{2}{\mu} \ e^{ \pi(\eh^+_1+\eh^+_2)} \ \frac{
{\CR_1 }(\Ep_{1}) {\CR_1 }(\Ep_{2})- {\CR_1 }(\Em_{1})
{\CR_1 }(\Em_{2})
}{\sinh{\pi(\eh_{1}^{+}-\eh_{1}^{-})}\sinh{\pi(\eh_{1}^{+}-\eh_{2}^{-})}}
\, . 
 \la{CRfinalh}} 
This expression can be safely analytically continued for $\eh>0$.
Note that the scattering amplitude (\ref{CRfinalh}) is of order
$1/\mu$ compared to the leading order, as it should.

We still need to show that the integral over the imaginary axis, $\e'
= i q$, vanishes.  The integral in question is
\eqn\Iint{ I = \int_{-\infty}^\infty dq \, F(q), \qquad F(q)= 4i\, 
\frac{\sin^{2}{2\pi q }\ \CR_1 (\kk+iq+\hf i)) \CR_1  ( \kk-iq+\hf
i)}{(\cosh{2\pi{\Delta}_{+}}+\cos{2\pi q})
(\cosh{2\pi{\Delta}_{-}}+\cos{2\pi q}) }.  } 
We split the interval into segments of length 1 and use the
quasi-periodicity of the integrand:
\eqn\Ibis{ I = \sum _{n\in \ZZ} \int _{-1/2}^{1/2} d q F(q+ n) = \int
_{-1/2}^{1/2} d q \sum _{n\in \ZZ} e^{i n \phi(q)} \, F(q)\, , \quad
e^{i\phi(q)}= \frac{e^{2\pi\eh} -e^{2\pi iq}}{ e^{2\pi\eh}-e^{-2\pi i
q}} \, .  } The sum of the unitary numbers gives $e^{2\pi\eh} \, \d(
q)$, so that indeed $I\sim F(0)=0$.

    \section{Calogero Hamiltonian for  general representations}
\label{app:canonical}

In this appendix, we derive the explicit form of the
Calogero Hamiltonian for arbitrary representation.
In particular, we focus on the second part of
the Hamiltonian (\ref{Hadj}) which is important
in the continuum limit,
\begin{equation}
H_1=\frac{1}{2}\sum_{j\neq k}^N \frac{\CD(E_{j}^{k})\CD(E_{k}^{j})}{
(x _j-x _k)^2}
\, .
\la{Hadj}
\end{equation}

In general, the wave function associated with the Young diagram whose
box (resp.  anti-box) part is described by $Y_1$ (resp.  $Y_2$) is
obtained by applying the Young symmetrizer associated with $Y_1$ to
the upper indices and the Young symmetrizer associated with $Y_2$ to
the lower indices and satisfies the tracesss condition for any pair of
the upper and lower indices.  More explicitly, if we write
\begin{equation}
\pi_Y=\sum_{\sigma \in S_n} \pi(Y, \sigma)\sigma
\end{equation}
as the Young symmetrizer for the Young diagram $Y$, the state
associated with the representation $Y_1, Y_2$ can be written in terms
of the adjoint wave functions as,
\begin{equation}\label{m_gwf}
\Psi^{j _1\cdots j_n}_{k_1\cdots k_n}(\X)=\sum_{\sigma,\tau\in S_n}
\pi(Y_1,\sigma) \pi(Y_2,\tau) f_1(\X)^{j _{\sigma(1)}}_{k_{\tau(1)}}
\cdots f_n(\X)^{j _{\sigma(n)}}_{k_{\tau(n)}}-\cdots\,.
\end{equation}
Here, again, $\cdots$ represents the terms which are needed 
to keep the traceless condition
for $\Psi^{j _1\cdots j_n}_{k_1\cdots k_n}(\X)$.

For a general representation (\ref{m_gwf}) we need to identify the
index sets $j_1,\cdots,j_n$ with $k_1,\cdots,k_n$.  In order to keep
track of the original index structure, it will be useful to represent
the state vector in the form
\begin{eqnarray}
&&\sum_{j _n, k_n}^N \Psi^{j _1\cdots j_n}_{k_1\cdots k_n}(\X)
|j_1\cdots j_n\rangle \langle k_1,\cdots, k_n|\nonumber\\
&& \quad \rightarrow \sum_{j _n}^N w_{j _1\cdots j_n}(x ) |j_1\cdots
j_n\rangle \langle j_1,\cdots, j_n|\\
&& w_{j _1\cdots j_n}(x )=
\Psi^{j _1\cdots j_n}_{j _1\cdots j_n}(x )
\end{eqnarray}
The element of the permutation on the upper (resp.  lower) indices can
be applied to the ket (resp.  bra) state as,
\begin{eqnarray}
&&\overrightarrow{\sigma}|j_1\cdots j_n\rangle \langle j_1,\cdots, j_n|
=|j_{\sigma(1)} \cdots j_{\sigma(n)} \rangle \langle j_1,\cdots, j_n|\,,\\
&&\overleftarrow{\sigma}|j_1\cdots j_n\rangle \langle j_1,\cdots, j_n|
=| j_1,\cdots, j_n \rangle \langle j_{\sigma(1)} \cdots j_{\sigma(n)} |\,.
\end{eqnarray}
Via such operators, one can define the projection into the irreducible
representations by the Young symmetrizer as (\ref{m_gwf}).

We are going  to calculate the action of the Hamiltonian to the
state of the representation $(Y_1,Y_2)$,
\begin{eqnarray}
\Psi^{Y_1}_{Y_2}(x )&=&\sum_{\sigma,\tau\in S_n} \pi(Y_1,\sigma) \pi(Y_2,\tau)
\sum_{j _1,\cdots, j_n}^N
w_{j _1,\cdots, j_n}(x )
\cdot\nonumber\\
&&~~~\cdot \overrightarrow{\sigma}\,\overleftarrow{\tau}\,
|j_1,\cdots,j_n\rangle \langle j_1\cdots, j_n|
\end{eqnarray}
The operator $\CD(E_{j k})$ in the interaction Hamiltonian
${H}_1$ takes the form
\begin{equation}
\CD(E_{j k})=\sum_{a=1}^n (\overrightarrow{E}^{(a)}_{j k}-
\overleftarrow{E}^{(a)}_{j k})
\end{equation}
where
\begin{eqnarray}
&& \overrightarrow{E}^{(a)}_{j k}|l_1\cdots l_n\rangle
\langle m_1,\cdots,m_n|=
 \delta_{k,l_a}|l_1,\cdots, j,\cdots, l_n\rangle
\langle m_1,\cdots,m_n|
\nonumber\\
&& \overleftarrow{E}^{(a)}_{j k}
|m_1,\cdots, m_n\rangle
\langle l_1\cdots l_n |=\delta_{j ,l_a}|m_1,\cdots, m_n\rangle
\langle l_1,\cdots, k,\cdots, l_n|
\end{eqnarray}
It is easy to check that\footnote{ We note that the operators
$\overleftarrow \sigma$ and $\overleftarrow E_{j k}^{(a)}$ act on the
state from the rightmost operator.  Namely $\overleftarrow\sigma \,
\overleftarrow\tau |l_1\cdots l_n\rangle \langle k_1,\cdots, k_n|
=|l_1\cdots l_n\rangle \langle k_{\sigma\tau(1)},\cdots,
k_{\sigma\tau(n)}|$.  }
\begin{equation}
\overrightarrow \sigma \cdot \overrightarrow E^{(a)}_{j k}=
\overrightarrow E^{(\sigma^{-1}\cdot b)}_{j k}\overrightarrow
\sigma\,, \qquad \overleftarrow \sigma \overleftarrow E^{(a)}_{j k}=
\overleftarrow E^{(\sigma^{-1}\cdot a)}_{j k}\overleftarrow \sigma
\end{equation}
and 
\begin{equation}
[\overrightarrow \sigma, H_1]=[ \overleftarrow \sigma, H_1]=0\,.
\end{equation}
Therefore the action of $\cH$ to $\Psi^{Y_1}_{Y_2}(x )$
can be written as,
\begin{eqnarray}
H_1\,\Psi^{Y_1}_{Y_2}(x )&=&
\sum_{\sigma,\tau\in S_n} \pi(Y_1,\sigma) \pi(Y_2,\tau)
\overrightarrow{\sigma}\,\overleftarrow{\tau}
H_1 w(x )
\nonumber\\
w(x )&=& \sum_{j _1,\cdots, j_n}^N w_{j _1,\cdots, j_n}(x
)|j_1,\cdots,j_n\rangle \langle j_1\cdots, j_n| \, .
\end{eqnarray}
When we evaluate the action of $H_1$ to the state $w(x )$, it is
useful to split it into the four parts,
\begin{eqnarray}
H_1&=& \hh_1+\hh_2+\hh_3+\hh_4\,,\\
\hh_1&=& \frac{1}{2}\sum_{j \neq k}^N \sum_{a=1}^n \frac{\rE^{(a)}_{j k}
\rE^{(a)}_{kj}+\lE^{(a)}_{j k}\lE^{(a)}_{kj}}{(x _j-x _k)^2}\\
\hh_2&=& -\sum_{j \neq k}^N \sum_{a=1}^n \frac{\rE_{j k}^{(a)}\lE_{kj}^{(a)}}{
(x _j-x _k )^2}\\
\hh_3&=& \frac{1}{2}\sum_{j \neq k}^N \sum_{a\neq b}^n \frac{\rE^{(a)}_{j k}
\rE^{(b)}_{kj}+\lE^{(a)}_{j k}\lE^{(b)}_{kj}}{(x _j-x _k )^2}\\
\hh_4&=& -\sum_{j \neq k}^N \sum_{a\neq b}^n \frac{\rE_{j k}^{(a)}\lE_{kj}^{(b)}}{
(x _j-x _k )^2}
\end{eqnarray}
The action of $\hh_1, ..., \hh_4$ to $w(x )$ can be evaluated as
\begin{eqnarray}
\hh_1 w(x )&=& \sum_{a}^n v_{j_a}(x ) w(x )
\,,\quad v_j (x)=\sum_{k(\neq j)}^N \frac{1}{(x _k -x _j)^2}\, ,\\
\hh_2 w(x )&=& -\sum_{j_1,\cdots,j_n}^N \left( \sum_{a}^n\sum_{j
}^N\frac{\delta(j\neq j_a)}{(x _j-x _{j_a})^2}
w_{j_1,\cdots,j,\cdots,j_n} \right)
|j_1\cdots j_n\rangle \langle j_1\cdots j_n|\, ,\\
\hh_3 w(x )&=& \frac{1}{2}\sum_{l_1,\cdots,c_n}^N \sum_{a\neq b}^n
\delta(j_a\neq j_b)\frac{w_{j_1,\cdots,j,\cdots,j_n}}{
(x _{j_a}-x _{j_b})^2}
 (\overrightarrow{(ab)}+\overleftarrow{(ab)})
 \cdot\nonumber\\
&&~~~~~~~~~~
\cdot
|j_1\cdots j_n\rangle \langle j_1\cdots j_n|\, ,\\
\hh_4 w(x )&=& -\sum_{a\neq b}^n\sum_{j_1,\cdots,j_n}^N
\frac{w_{j_1,\cdots,j_a,\cdots,j_b,\cdots,j_n}}{(x _{j_b}-x _{j_a})^2}
\overleftarrow{(ab)}|j_1\cdots j_n\rangle \langle j_1\cdots j_n|\,.
\end{eqnarray}
Here $(ab)$ represents the transposition and 
$\delta(a\neq b)=1-\delta_{ab}$.
After the (anti-)symmetrization by Young symmetrizer, 
the action of $\cH_1$ becomes,
\begin{eqnarray}
H_1&=&H^{\mathrm{(free)}} +V \\
(H^{\mathrm{(free)}}w)_{j_1,\cdots,j_n}&=& \sum_{a}^n \left(v(j_a)
w_{j_1,\cdots, j_n} -\sum_{j }^N\frac{\delta(j\neq j_a)}{(x _j-x
_{j_a})^2} w_{j_1,\cdots,j,\cdots,j_n} \right)\\
(V w)_{j_1,\cdots,j_n}&=&\frac{1}{2}\sum_{a\neq
b}^L\frac{\delta(j_a\neq j_b)}{(x _{j_a}- x
_{j_b})^2}(q^L_{ab}+q^R_{ab})w_{j_1,\cdots, j_a , \cdots, j_b, \cdots,
j_n}\nn\\
&&-\sum_{a\neq b}^n\delta(j_a\neq j_b) \, q^R_{ab}
\frac{w_{j_1,\cdots,j_ b,\cdots,j_b,\cdots,j_n}}{
(x _{l_n}-x _{l_m})^2}\, .
\end{eqnarray}
Here $q^{L,R}_{ab}$ is the eigenvalues of $\overrightarrow{(ab)},
\overleftarrow{(ab)}$ after the projection by Young symmetrizer.
It depends on the location of $a,b$ in the Young tableau $Y_1,Y_2$.

The piece $\cH^{\mathrm{(free)}}$ is a direct sum of $\cH_1$ for the
adjoint representation for the indices $j_1,\cdots,j_n$.  It
represents $n$ non-interacting quasiprticles.  In particular, this
part depends only on the number of boxes $|Y_1|=|Y_2|=n$

On the other hand, $\cV$ represents the interaction between the
quasiparticles.  It depends on the representations $(Y_1,Y_2)$ and is
somehow complicated.  The interaction simplified for $A_n$, $B_n$ and
$C_n$ representations.  The interaction $\cV$ among the tips of the
folded string becomes,
\begin{eqnarray}
A_L&\, :\,& V=0\,,\\
B_L &\, :\,& (V w)_{i_1,\cdots,i_L}=
\sum_{n\neq m}^L\frac{\delta(i_n\neq i_m)}{(x_{i_n}-
x_{i_m})^2}(w_{i_1,\cdots,c_L}
-w_{i_1,\cdots,i_m,\cdots,i_m,\cdots,i_L})\,,\\
C_L&\, :\,&(V w)_{i_1,\cdots,i_L}=-
\sum_{n\neq m}^L\frac{\delta(i_n\neq i_m)}{(x _{i_n}-
x _{i_m})^2}w_{i_1,\cdots, i_L}\,.
\end{eqnarray}

\end{document}